 \documentclass[10pt,preprint]{aastex}  
 \usepackage{natbib}
\def\ang{\AA}

\def\gapprox{\lower.4ex\hbox{$\;\buildrel >\over{\scriptstyle\sim}\;$}}
\def\lapprox{\lower.4ex\hbox{$\;\buildrel <\over{\scriptstyle\sim}\;$}}

\shortauthors{ASCHWANDEN 2016}
\shorttitle{Refined Energetics of CMEs}

\begin{document}

\title{         Global Energetics of Solar Flares: 
		VI. Refined Energetics of Coronal Mass Ejections }

\author{        Markus J. Aschwanden$^1$}

\affil{		$^1)$ Lockheed Martin, 
		Solar and Astrophysics Laboratory, 
                Org. A021S, Bldg.~252, 3251 Hanover St.,
                Palo Alto, CA 94304, USA;
                e-mail: aschwanden@lmsal.com}

\begin{abstract}
In this study we refine a CME model presented in an earlier study
on the global energetics of solar flares and associated CMEs, and
apply it to all (860) GOES M- and X-class flare events observed
during the first 7 years (2010-2016) of the Solar Dynamics Observatory
(SDO) mission, which doubles the statistics of the earlier study.
The model refinements include: (1) the CME geometry in terms of a
3D sphere undergoing self-similar adiabatic expansion; (2) the
inclusion of solar gravitational deceleration during the acceleration
and propagation of the CME, which discriminates eruptive and confined 
CMEs; (4) a self-consistent relationship between the CME center-of-mass 
motion detected during EUV dimming and the leading-edge motion observed 
in white-light coronagraphs; 
(5) the equi-partition of the CME kinetic and thermal energy; and 
(6) the Rosner-Tucker-Vaiana (RTV) scaling law. The refined CME model
is entirely based on EUV dimming observations (using AIA/SDO data) and
complements the traditional white-light scattering model (using 
LASCO/SOHO data), and both models are independently capable to 
determine fundamental CME parameters such as the CME mass, speed,
and energy. Comparing the two methods we find that: (1) LASCO is
less sensitive than AIA in detecting CMEs (in 24\% of the cases); 
(2) CME masses below $m_{cme} \lapprox 10^{14}$ g are under-estimated
by LASCO; (3) AIA and LASCO masses, speeds, and energy agree closely in the
statistical mean after elimination of outliers; (4) the CMEs parameters
of the speed $v$, emission measure-weighted flare peak temperature $T_e$,
and length scale $L$ are consistent with the following scaling laws 
(derived from first principles): $v \propto T_e^{1/2}$, $v
\propto (m_{cme})^{1/4}$, and $m_{cme} \propto L^2$.
\end{abstract}

\section{		    INTRODUCTION			}

There exist over 2000 refereed publications on the phenomenon of 
{\sl Coronal Mass Ejections (CMEs)}, and at least 80 review articles, 
such as, for instance, Schwenn (2006), Chen (2011), Webb and Howard 
(2012), or Gopalswamy (2016).
A deeper understanding of the physical processes that occur during
a CME can be gained by infering physical scaling laws and
statistical distributions, which both require ample statistics.
Most CME studies focus on a single or on a small number of events, 
while statistical studies are rare. We identified about 60 studies
that contain large statistics ($\approx 10^2-10^5$ events) of observed 
and physical CME parameters, such as the sizes and locations of CMEs 
(Hundhausen 1993; Bewsher et al.~2008; Wang et al.~2011), 
the CME speed, acceleration, mass, and energy (Moon et al.~2002; 
Yurchyshyn et al.~2005; Zhang and Dere 2006; Cheng et al.~2010; 
Bein et al.~2011; Joshi and Srivastava 2011; Gao et al.~2011),
and the associated flare hard X-ray fluxes, 
fluences, and durations (Yashiro et al.~2006; Aarnio et al.~2011).
The most extensive statistics of CME parameters is provided in
on-line catalogs of CME events detected with the white-light method,
mostly from the {\sl Large-Angle and Spectrometric Coronagraph Experiment 
(LASCO)} onboard the {\sl Solar and Heliospheric Observatory (SOHO)} 
(Brueckner et al.~1995), such as the CDAW, Cactus, SEEDS, and CORIMP
catalogs of CME events, but also from STEREO/COR2 (see web links in 
Section 4.1).

A dedicated effort has been undertaken to study the global energetics
and energy partition of solar flares and associated CME events, 
which yields statistics of physical CME parameters and provides 
tests of the underlying physical scaling laws. The analyzed data sets 
include all M and X-class flares during the {\sl Solar Dynamics 
Observatory (SDO)} mission (Pesnell et al.~2011). In the previous
studies we measured the various types of energies that can be 
detected during flares and CME events, including the dissipated 
magnetic energy (Aschwanden Xu, and Jing 2014; Paper I), 
the multi-thermal energy (Aschwanden et al.~2015; Paper II), 
the nonthermal energy (Aschwanden et al.~2016; Paper III), 
the kinetic and gravitational energy of associated CMEs 
(Aschwanden 2016; Paper IV), and the energy closure
(Aschwanden et al.~2017; Paper V). Regarding CME energetics,
there is the traditional white-light scattering method on one
side, and the more novel EUV-dimming method on the other side,
which both will be extensively discussed in this paper
(for references of both methods see Paper IV and references therein).

In this study (Paper VI) we refine the CME model presented in 
Paper IV in a number of ways, which entails larger statistics,
the discrimination between eruptive and confined CMEs,
the deceleration caused by the gravitational potential,
a self-consistent relationship between the center-of-mass motion
and the leading-edge motion, the equi-partition of the kinetic and 
thermal energy in CMEs, and the Rosner-Tucker-Vaiana (RTV) scaling law
(Rosner et al.~1978).
The intent of this study is the derivation of more accurate values of 
the CME mass, speed, and energy than in previous work. Moreover, this 
refined kinematic CME model provides physical CME parameters from EUV 
dimming data alone, which complements the traditional method using 
white-light coronagraphic observations. 

The content of this paper contains an analytical description and
derivation of the refined CME model from first principles (Section 2),
observations from the {\sl Atmospheric Imager Assembly (AIA)}
(Lemen et al.~2012) onboard the {\sl Solar Dynamics Observatory (SDO)}
(Pesnell et al.~2011) and data analysis in terms of forward-fitting
the refined CME model to the EUV dimming data (Section 3), 
discussion of the CME measurements in the context of previous work
(Section 4), and conclusions (Section 5). 

\section{		ANALYTICAL MODEL  			}

In a previous study on the global energetics of flares and CMEs
we used a forward-fitting method of a parameterized CME model to fit 
the time evolution of EUV dimming (Aschwanden 2016; Paper IV). In the new
study presented here we refine the method of CME modeling by including 
additional effects, such as: 
(1) The discrimination between eruptive and confined CMEs;
(2) the deceleration caused by the gravitational potential;
(3) a self-consistent relationship between the center-of-mass and the 
leading edge motion observed in white-light coronagraphs; 
(4) the equi-partition of the CME kinetic and thermal energy, and 
(5) the Rosner-Tucker-Vaiana (RTV) law. The refined method allows 
us to model both confined and eruptive flares with the same model, and
to derive more reliable values of the CME mass, speed, and energy than
in previous work. Moreover, this refined kinematic CME model provides
physical CME parameters from EUV dimming data alone, which complements 
the traditional method using white-light coronagraphic observations. 

\subsection{		The CME Geometry				}

We start with a geometric model of the time-dependent CME volume as
depicted in Fig.~1. Before the launch of the CME, all the plasma that
will feed the later expanding CME volume is confined in a volume $V_0$
with an area $A=L^2$ (with an unprojected length scale $L$) on the 
solar surface, within a vertical height extent $h_0$ that corresponds to  
the temperature-dependent electron density scale height $\lambda_{n_e}(T_e)$
observed at the beginning of the flare at temperature 
$T_0 = T_e(t=t_{start})$,  
\begin{equation}
	V_0 = L^2 h_0 = L^2 \ \lambda_{n_e}(T_0) \ ,
\end{equation}
where the electron density scale height $\lambda_{n_e}$ is,
\begin{equation}
	\lambda_{n_e}(T_0) =  
	{2\ k_B\ T_0 \over \mu\ m_H\ g_{\odot}}
	\approx 4.7 \times 10^9 \ 
	\left( {T_0 \over 1\ {\rm MK}} \right) \quad [cm] \ .
\end{equation}
where $k_B=1.38 \times 10^{-16}$ erg K$^{-1}$ is the Boltzmann constant, 
$\mu \approx 1.27$ is the mean molecular weight,
$m_H =1.67 \times 10^{-24}$ g is the hydrogen mass, and 
$g_{\odot}=\Gamma M_{\odot}/R_{\odot}^2 =2.74 \times 10^4$ cm s$^{-2}$  
is the solar gravity acceleration. 
This initially surface-aligned volume $V_0$ is shown from a top-down 
view for an equatorial CME at a longitudinal angle $\rho = l$ 
(between the observer's line-of-sight and the CME center-of-mass
trajectory) from 
the central meridian at time $t_1$ in Fig.~1 (top panel). For an 
arbitrary CME launch position the angle $\rho$ can be calculated 
from the heliographic position at longitude $l$ and latitude $b$
from spherical trigonometry,
\begin{equation}
	\cos{(\rho)} = \cos{(l)} \cos{(b)} \ .
\end{equation}
The top side of the pre-lauch CME volume at an altitude $h_0$ above the 
photosphere has a distance of
\begin{equation}
	r_0 = R_{\odot} + h_0 \ ,
\end{equation}
from Sun center, where $R_{\odot}$ is the solar radius, and the projected 
distance $x_0$ from Sun center in the plane-of-sky is (Fig.~1, bottom panel), 
\begin{equation}
	x_0 = (R_{\odot} + h_0) \sin{(\rho)} \ .
\end{equation}
At the launch time $t = t_1$ of the CME, the
plasma confined in the original volume $V_0=v(t \le t_1)$ will start to stream 
into a spherical volume that expands subsequently (as shown for 10 time 
steps from $t_1$ to $t_{10}$ in Fig.~1, top panel), where the bottom of 
the sphere stays connected at a coronal height $h_0$, and the CME center of mass
moves in radial direction away from the Sun at the radial position $r(t)$,
with a CME radius of $R(t)$ (Fig.~1, bottom panel), 
\begin{equation}
	R(t) = r(t) - r_0  \ ,
\end{equation}
The projected position $x(t)$ of the CME center of mass is
\begin{equation}
	x(t) = r(t) \ \sin{(\rho)} \ ,
\end{equation}
and the projected position $x_{LE}(t)$ of the CME leading edge is
\begin{equation}
	x_{LE}(t) = x(t) + R(t) \ .
\end{equation}
The projected leading edge position $x_{LE}(t)$ is important for relating
the timing of the coronagraphic CME detection to the EUV dimming
model. If a coronagraph detects a CME at a projected location
$x_{LE}(t)$ at time $t$, it follows from Eqs.~(6-8) that the radial
distance $r(t)$ of the CME center-of-mass from Sun center is,
\begin{equation}
	r(t) = {r_0 + x_{LE}(t) \over 1 + \sin{(\rho)} } \ .
\end{equation}

Thus the geometric CME model can be described with the
time evolution of the radial distance $r(t)$ of the CME center-of-mass
position from Sun center, the heliographic position $(l,b)$ of the 
flare (or CME launch) site, the (unprojected) length scale
$L$ of the CME footpoint area, and with the emission 
measure-weighted temperature $T_w$ (which defines $h_0$, $r_0$, 
and $V_0$). The emission measure-weighted temperature $T_w$ 
can be determined from the differential emission measure distributions 
$dEM(T)/dT$ at the peak time of a flare,
\begin{equation}
	T_w	= {\int T \ [dEM(T)/dT]\ dT \over \int [dEM(T)/dT]\ dT} \ ,
\end{equation}
as described in Papers II and IV. 

\subsection{	The CME Acceleration Phase			}
	
We define the start time $t_1$ of the CME acceleration phase by
the time when the total emission measure profile $EM(t)$ peaks
during a flare time interval. This follows from the fact that
the time evolution of the EUV dimming has only a physical meaning 
when it decreases, which requires a temporal peak at the beginning 
of the EUV dimming phase. Thus the initial parameters at the
beginning of the EUV dimming phase or before $(t \le t_1)$ are
according to the geometric model depicted in Fig.~1,
\begin{equation}
	(t \le t_1)
        \left\{ \begin{array}{ll}
        a(t)    & = 0		\\
        v(t)    & = 0		\\
        r(t)    & = h_0		\\
        R(t)    & = 0		\\
        x(t)    & = x_0		\\
        \end{array}
        \right.
\end{equation}
where $a(t)$ is the acceleration and $v(t)$ the velocity of the
CME center of mass.
We employ the simplest model for the CME acceleration phase, namely
a constant acceleration $a$ during a time interval $\tau_{acc}$
(similar to Paper IV), i.e., $t_1 \le t \le t_2$, defining an acceleration 
end time $t_2$ of 
\begin{equation}
	t_2 = t_1 + \tau_{acc} \ .
\end{equation}
The time evolution of the dynamical parameters during this acceleration
phase $t_1 \le t \le t_2$ is thus, 
\begin{equation}
	(t_1 \le t \le t_2)
        \left\{ \begin{array}{ll}
        a(t)    & = a_1		\\
        v(t)    & = a_1 (t - t_1) \\
        R(t)    & = (a_1/2) (t - t_1)^2 \\
        r(t)    & = r_0 + (a_1/2) (t - t_1)^2 \\
        \end{array}
        \right.
\end{equation}
while the projected positions $x(t)$ and $x_{LE}(t) $ follow from Eqs.~(7-8).
At the end time $t_2$ of the acceleration phase, the values of the
kinematic parameters are,
\begin{equation}
	(t = t_2)
        \left\{ \begin{array}{ll}
        a_2 = a(t=t_2)  & = a_1		\\
        v_2 = v(t=t_2)  & = a_1 (t_2 - t_1) = a_1 \tau_{acc} \\
        R_2 = R(t=t_2)  & = (a_1/2) (t_2 - t_1)^2 = (a_1/2) \tau_{acc}^2 \\
        r_2 = r(t=t_2)  & = r_0 + (a_1/2) \tau_{acc}^2 \\
        \end{array}
        \right.
\end{equation}
A graphic representation of the evolution of the acceleration $a(t)$, 
speed $v(t)$, and position $r(t)$ of the CME center of mass is shown
in Fig.~2.

\subsection{	Gravitational Deceleration Phase			}

We define now a third time interval, $t > t_2$, when the CME is not
accelerated any further, but is only subject to the deceleration $a(t)$
caused by the gravity force, which is important for CMEs that reach a
speed near or below the escape speed from the Sun,
\begin{equation}
	a(t) = - {\Gamma M_\odot \over r(t)^2} \ ,
\end{equation}
where $\Gamma$ is Newton's gravitational constant, and $M_{\odot}
= 2 \times 10^{33}$ g is the solar mass. We see that this equation 
represents a second-order differential equation of the type 
$(r^{\prime \prime} r^2 = const)$ that has no simple analytical solution. 
However, since the velocity becomes almost constant after the
acceleration phase, we can approximate the time dependence of
the radial distance with $r(t) \approx r_2 + v_2 (t-t_2)$ to first
order, which leads to the following time evolution of the gravitational
deceleration,
\begin{equation}
	a(t) \approx - \Gamma M_\odot [r_2 + v_2 (t-t_2)]^{-2} \ ,
\end{equation}
expressed as an explicit function of time. We can now calculate
the time evolution of the CME center-of-mass speed straightforwardly
by time integration of the deceleration (Eq.~16),
\begin{equation}
	v(t) = v_2 + \int_{t_2}^t a(t) \ dt =
	v_2 - {\Gamma M_\odot \over v_2}
	\left\{ {1 \over r_2} - {1 \over r_2 + v_2 (t-t_2)} \right\} \ .
\end{equation}
We see that the speed $v(t)$ is monotonically decreasing with time
after $t > t_2$, and converges asymptotically to the final speed 
$v_{\infty}$ (at time $t \mapsto \infty$), 
\begin{equation}
	v_{\infty} = v(t=\infty) = v_2 - {\Gamma M_\odot \over v_2\ r_2} \ .
\end{equation}
Finally we can also calculate a more accurate value for the
evolution of the radius $R(t)$ of the CME 
by time integration of the velocity $v(t)$ given in Eq.~(17),
using the integral $\int r^{-1}\ dr = \ln{(r)}$,
\begin{equation}
	R(t) = R_2 + \int_{t_2}^t v(t) \ dt =
	R_2 + [v_2 - {\Gamma M_{\odot} \over v_2 r_2}] (t - t_2)
	+ {\Gamma M_\odot \over (v_2)^2}
	\left\{ \ln{[r_2 + v_2 (t-t_2)]} - \ln{[r_2]} \right\} \ .
\end{equation}
The distance $r(t)$ of the CME center-of-mass from Sun center
follows then (with Eq.~6),
\begin{equation}
	r(t) = r_0 + R(t) \ ,
\end{equation}
and the projected position $x(t)$ of the CME center-of-mass
and $x_{LE}(t)$ of the CME leading edge follow from Eqs.~(7-8).

\subsection{	Confined and Escaping CMEs		}

The fate of whether the expanding CME sphere escapes during the
eruption from the Sun, or whether it turns into a stalled 
(failed) eruption that comes to a halt and falls back to the Sun,
depends on whether the CME reaches the critical escape velocity
during the initial acceleration phase or not. Since our dynamic
model is designed to reach a maximum speed $v_2$ at the end of
the acceleration phase at $t=t_2$, which is $v_2=a_1 \tau_{acc}$
(Eq.~14), the critical escape speed $v_{CME}$ has to be calculated
at this position $r_2=r_0+(a_1/2) \tau_{acc}^2$ (Eq.~14), which
follows from Eq.~(18) by setting $v_\infty=0$, yielding with 
Eqs.~(4) and (13),
\begin{equation}
	v_{esc}(r= r_2) = \left( {2 \Gamma M_{\odot} \over r_2 } \right)^{1/2}
	= \left( {2 \Gamma M_{\odot} \over R_{\odot} + h_0 + (a_1/2)\tau_{acc}^2} 
	\right)^{1/2} \ .
\end{equation}
Thus, our analytical model describes the time evolution of both
a confined flare, if $v_2 < v_{esc}(r_2)$, and an eruptive CME,
if $v_2 \ge v_{esc}(r_2)$.   	

\subsection{	Coronagraphic CME Detections			}

It is also useful to relate the CME velocity $v$ to the propagation
distance $r$, which allows us to compare the speeds of our model
with coronagraphic observations. From the conservation of kinetic
energy and the gravitational potential at distances $r$ and $r_2$, 
\begin{equation}
	{1\over 2} m_{cme} v^2 - {1\over 2} m_{cme} (v_2)^2 =
	+ {\Gamma M_{\odot} m_{cme} \over r}  
	- {\Gamma M_{\odot} m_{cme} \over r_2} \ , 
\end{equation}
we can obtain the velocity $v[r(t)]$ at any location $r(t)$ after the
acceleration phase, at $t \ge t_2$,
\begin{equation}	
	v[r(t)] = \left[ (v_2)^2 - 2 \Gamma M_{\odot}
		\left({1 \over r_2} - {1 \over r(t)}\right) \right]^{1/2} \ .
\end{equation}
For instance, if a CME is detected with a coronagraph at time
$t=t_3$ at a distance $x_3$ given by the occulting disk, which 
corresponds to a radial distance $r_3=(r_0 + x_3)/(1 + \sin{(\rho))}$
from Sun center according to Eq.~(9), we can predict the velocity 
$v_3$ at this particular location and time,  
\begin{equation}	
	v_3    = \left[ (v_2)^2 - 2 \Gamma M_{\odot}
		\left( {1 \over r_2} - {1 \over r_3} \right) \right]^{1/2} \ .
\end{equation}
For LASCO observations for instance, the occulting disk is at
$x_3 = x_{LASCO} \approx 4 R_{\odot}$, where the CME mass $m_{LASCO}$,
the speed $v_{LASCO}$, and the detection time $t_{LASCO}$ is measured,
which can then be compared with the values $m_3, v_3$, and $t_3$
of our CME model. The predicted time $t_3$ of CME detection with
LASCO can be computed in our CME model from the parameters
$(r_2, v_2, v_3)$ using Eq.~(17),
\begin{equation}
	t_3 = t_2 + \left( {1 \over v_2} \right)
		\left[ \left( {(v_3-v_2) v_2 \over \Gamma M_{\odot}} + {1 \over r_2}
		\right)^{-1} - r_2 \right]\ .
\end{equation}

\subsection{	Adiabatic CME Expansion and EUV Dimming }

Our dynamic CME model can be fitted to data that measure the
EUV dimming, which requires the time evolution of the total
emission measure $EM(t)$. In our simple CME model we assume
a purley adiabatic expansion, where no energy is exchanged
across the boundaries of a CME, and thus predicts that the
average electron density changes reciprocally to the expanding
CME volume (Paper IV), in order to conserve the number of particles,
\begin{equation}
	q_{EM}(t) = {EM(t) \over EM_0} = {V_0 \over V(t) } \ ,
\end{equation}
where $EM_0$ and $V_0$ are the initial total emission measure
and the initial volume (as defined in Eq.~1). Therefore, all
that is needed to calculate the time evolution of the EUV dimming
$EM(t)$ is the time-dependent volume $V(t)$, which we define as
the sum of the coronal source volume $V_0$ and the spherically
expanding CME volume,
\begin{equation}
	V(t) = V_0 + {4 \over 3} \pi R(t) \ ,
\end{equation}	
where the CME radius $R(t)$ is defined with Eq.~(13) during the
acceleration phase, and with Eq.~(19) after the acceleration phase,
during gravitational deceleration. 

Since there is always some background emission measure  
observed in every flare and CME, originating from the non-flaring
part of the Sun, we have to correct the 
observed emission measure by the fraction of the background 
component $q^{bg}$,
\begin{equation}
	EM^{fit}(t) = EM^{max} \left[ q^{bg} + (q^{peak}-q^{bg}) 
		q_{EM}(t) \right] \ ,
\end{equation}
where $EM^{max}$ is the maximum of the observed emission measure
and $q^{peak}$ is the fraction of the modeled peak emission measure 
to the absolute maximum of the observed total emission measure. 
We see from Eq.~(28) that the modeled emission measure has the initial value
$EM^{fit}(t=t_1)=EM^{max} q^{peak}$ (for $q_{EM}(t=t_1)=1$), and
asymptotically approaches the value 
$EM^{fit}(t=\infty)=EM^{max} q^{bg}$ (for $q_{EM}(t=\infty) \mapsto 0$).

\subsection{	Energy Equi-Partition Model			}

The observed EUV dimming exhibits the fastest change in the
initial phase of the CME expansion (in the lower corona), 
while the later expansion in the heliosphere causes very
small changes that asymptotically reach unmeasurable small values.
The final CME speed is therefore very weakly constrained by 
the emission measure profile $EM(t)$. It is therefore desirable
to test the final CME expansion speed $v(t)$ by other means,
for instance by making use of the assumption of energy partition 
between the kinetic and thermal energy contained in
the corresponding flare, which has been empirically found to be
closely fulfilled in a previous statistical study (Paper V),
\begin{equation}
	E_{kin} = {1 \over 2} m_{cme} v^2 
	\approx E_{th} = 3 n_e k_B T_w V \ .
\end{equation}
Since both forms of energy contain the volume-integrated CME mass,
$m_{cme} = n_e m_p V$, both the density and the volume cancel out, 
and yields a very simple relationship between the CME velocity
$v$ and the (emission measure-weighted) flare temperature $T_w$. 
If we apply this energy equi-partition to the maximum CME kinetic 
energy, which happens at $v_2=v(t=t_2)$ in our model, we obtain
\begin{equation}
	v_{2T} = \sqrt{ \left( {6 k_b \over m_p} \right) \ T_w }
	    = c_1 \ \left( {T_w \over 10\ {\rm MK}} \right)^{1/2}
	\qquad c_1 = 704\ {\rm km}\ {\rm s}^{-1} \ .
\end{equation}  
where we denote the velocity as $v_{2T}$ to indicate the temperature
model. From our previous study we measured (emission 
measure-weighted) flare temperatures in the range of $T_w = 3-14$ MK,
for which the energy equi-partition model (Eq.~30) predicts maximum
velocities in the range of $v_2=384-830$ km s$^{-1}$. 

\subsection{	The Rosner-Tucker-Vaiana Scaling Law		}

A scaling law for the energy balance between the heating rate and the
(conductive and radiative) cooling rate has been derived for quiescent
coronal loops (Rosner et al.~1978), which applies also to the turnover
point between the dominant heating phase and the dominant cooling phase
in solar flares
(Aschwanden and Tsiklauri 2009). The original formulation by Rosner
et al.~(1978) states the relationship $T_{max} \approx 1400 (p_0 L)^{1/3}$
between a loop apex temperature $T_{max}$, the approximately constant
loop pressure $p_0$, and the half length $L$ of a semi-circular loop.
Inserting the iso-thermal pressure of an ideal gas, $p = 2 k_B n_e T_e$,
yields then a relationship for the electron density at the footpoints
of flare loops (Aschwanden and Shimizu 2013, Eq.~14 therein),
\begin{equation}
	n_e = n_{e0} \ \left({ T_e \over 10\ {\rm MK}} \right)^2
	               \left({ L \over 10^{10}\ {\rm cm}} \right)^{-1} 
	\qquad n_{e0} = 8.4 \times 10^9 \ {\rm cm}^{-3} \ .
\end{equation}
The geometry of a post-flare arcade is typically a sequence of
semi-circular flare loops, which can be represented by a volume
that covers the flare or CME footprint area $A=L^2$ and has a
typical filling factor of $q_{fill} \approx 0.05$ for the
Euclidean volume $V=L^3$ (Aschwanden and Aschwanden 2008b), 
\begin{equation}
	V = V_0 \left({L \over 10^{10}\ cm}\right)^3
	        \left({q_{fill} \over 0.05}\right) \ ,
	\qquad V_0 = 0.5 \times 10^{29} \ {\rm cm}^{-3} \ .
\end{equation}
The total CME mass (initially confined in the flare volume) is then,
\begin{equation}
	m_{cme} = n_e\ m_p\ V_0 = n_e\ m_p\ L^2 h_0 \ ,
\end{equation}
where $m_p$ is the proton mass. Combining the Eqs.~(31-33), together
with the relationship of the equi-partition between the kinetic and
thermal energy (Eq.~30) we obtain then a total mass of 
\begin{equation}
	\left({ m_{cme} \over 10^{15}\ {\rm g}}\right) = 0.7 \ 
	    \left( {T_e \over 10\ {\rm MK}} \right)^2
	    \left({L \over 10^{10}\ cm}\right)^2
	    \left({q_{fill} \over 0.05}\right) \ .
\end{equation}
Expressing the CME velocity $v_{CME}=v_{2m}$ explicitly, we obtain 
the scaling law
\begin{equation}
	v_{2m} = 1.09\ c_1 
         \left( { m_{cme} \over 10^{15}\ {\rm g}} \right)^{1/4}
	 \left( { L \over 10^{10}\ {\rm cm} }  \right)^{-1/2}
	 \left( { q_{fill} \over 0.05 }  \right)^{-1/4}
	 \approx c_2 
	 \left({ m_{cme} \over 10^{15}\ {\rm g}}\right)^{(1/4)} \ , 
	 \qquad c_2 = 767\ {\rm km}\ {\rm s}^{-1} \ ,
\end{equation}
where we denote the velocity with $v_{2m}$ to indicate the mass model.
If we ignore the dependence on the loop length $L$ and the filling 
factor, we can approximately predict the maximum CME speed $v_2$ 
from the CME mass $m_{cme}$ alone. In our
analyzed data set we find CME masses of $m_{cme} \approx (0.01-40)
\times 10^{15}$ g, from which Eq.~(35) predicts CME speeds in the range
of $v_2 \approx 200-1600$ km s$^{-1}$. 

We have now two equi-valent relationships for the CME speed,
one that depends only on the temperature, i.e., 
$v_{2T} = (T_e/10$ MK)$^{1/2}$ $\times 704 $ km s$^{-1}$ (Eq.~30),
and one that depends only on the CME mass, i.e., 
$v_{2m} = (m_{cme}/10^{15}$\ g$)^{1/4} \times 767$ km s$^{-1}$ (Eq.~35).
The difference between the two methods provides an estimate of
systematic uncertainties. In our data analysis method we determine 
the maximum CME speed $v_2$ independently from the EUV dimming alone,
without making the assumption of energy equi-partition or tht RTV law,
but can use those assumptions as an additional test, besides 
comparisons of the speeds $v_2$ measured in white-light data
from LASCO.

\section{	OBSERVATIONS AND DATA ANALYSIS 		}

In a previous study (Aschwanden 2016; Paper IV),
the kinematic parameters of CMEs have been modeled from the EUV
dimming observed with AIA/SDO, and have been compared 
with the same parameters obtained from white-light data observed 
with LASCO/SOHO. In this study we present a refined CME model 
with extended data analysis, incorporating a number of additional 
effects (as enumerated at the beginning of Section 2) 
that were not taken into account in the previous study. 
The key parameters that we are interested in 
here are the CME mass, speed, and kinetic energy.

The analysis procedure is depicted in Fig.~3,
which consists of the measurements of the fluxes in each AIA wavelength
(Fig.~3a) and the DEM inversion of the total emission measure profile 
$EM(t)$ (Fig.~3e), which is
fitted with the theoretical CME model (red profile in Fig.~3e) described 
in Fig.~2, yielding the CME motion at a 
projected distance $x(t)$ (Fig.~3d), the velocity profile $v(t)$ (Fig.~3c), 
and the acceleration profile $a(t)$ (Fig.~3b). In the following we
describe the statistical distributions of the measured and best-fit
physical CME parameters, displayed in form of power law distributions
(Fig.~4) or Gaussian normal distributions (Fig.~5), from which we list
the parameter ranges, medians, power law slopes, and means and standard
deviations in Table 1.

\subsection{	AIA Observations 				}

The analyzed data set includes all GOES M- and X-class flare events 
recorded during the first 7 years (June 2010 - Nov 2016) of the SDO 
mission, which amounts to 864 events, where we doubled the previously 
analyzed data set from the first 3.5 years of the SDO mission (Paper I-V). 
Only 4 events of the AIA data set contained data gaps during the flare 
time interval, which are discarded
here, while 860 events remain for further analysis. We are using the same 
event numbering list as in Papers I-V, so the event numbers \#1, ..., \#399
are identical with the previous analysis, while the events \#400,
..., \#864 are new.
The time evolution of the analyzed images is subdivided into steps of 
$\Delta t=2$ min, covering the entire flare duration as defined by 
the start and end times (prolonged by a margin of 30 min) from 
the GOES flare list. 

\subsection{	Flare Temperatures 				}

The first step in our analysis of AIA flare data is the automated
differential emission measure (DEM) analysis, using the spatial-synthesis 
DEM code (Aschwanden et al al.~2013), which uses the 6 coronal
AIA wavelengths (94, 131, 171, 193, 211, 335 \ang ) and yields a
time sequence of DEM distributions $dEM(T,t)/dT$, as described in
the previous Papers II and IV. From these DEMs we determine the 
emission measure-weighted flare temperatures $T_w(t)$ (Eq.~10)
for each time step, and take then the maximum value 
$T_w=T_w(t=t_{max})$ of the time sequence to characterize
the (thermal) density scale height near the flare peak.
The distribution of these flare temperatures $T_w$ is shown
in Fig.~5a, which covers a range of $T_w^{max}=2.3-17.4$ MK and
has a mean of $T_w^{max}=9.3\pm2.4$ MK. Note that this flare 
temperature range is substantially higher than the pre-CME 
temperatures $T_e=1.2-5.0$ MK determined in the previous study 
(Paper II), which define the CME volume and mass (Eqs.~1, 33).

\subsection{	CME Source Parameters  				}

The source volume $V_0$ of the CME at the beginning of (or before) 
the expansion is defined in terms of the unprojected source area
$A$ and the vertical height $h_0$ (Eq.~1). The measurement of
the unprojected length scale $L$ is described in Section 2.3
in Paper IV, for which we find a range of $L=18-361$ Mm (Fig.~4a). 
The resulting CME footpoint or dimming area $A$ is simply defined 
as $A=L^2$ (Fig.~4b), where $L$ is the deprojected length scale.  

The preflare temperature $T_0$ defines then the emission measure 
scale height $h_0=\lambda_{n_e}(T_0)$ according to Eq.~(2). The pre-flare
temperature is measured at the start time $t_{start}$ of the GOES flare,
which amounts to an average value of $T_0=1.8\pm0.3$ MK
(with a scale height $\lambda_{n_e} \approx 0.12\ R_{\odot}$). 
The peak of the flare temperature is found at a mean value of
$T_w=9.3\pm2.4$ MK (with a scale height $\lambda_{n_e}=0.63\ R_{\odot}$). 
The temperature at the time $t_1$ of the CME launch, defined by the
start of the EUV dimming or peak value of the total emission measure,
is generally between the GOES flare start $t_{start}$ and the 
temperature maximum time $t_m=t(T=T_m)$, i.e., $t_0 < t_1 < t_m$,
at an average temperature of $T_m=4.7\pm1.2$ MK, with a
(with a scale height $\lambda_{n_e} = h_0 \approx  
0.31 \pm 0.08) R_{\odot}$ (Fig.~5b). The resulting CME 
source volumes (Eq.~1) of the analyzed 860 events vary in the range of 
$V_0=(0.0005-0.5)\times 10^{30}$ cm$^{3}$ (Fig.~4c).

The maximum emission measure per area $EM(t)/A$ displays the most
extended power law distribution (over two decades), with a power law
slope of $p=-2.6 \pm 0.6$ (Fig.~4d). The total emission measure is
defined in terms of the spatially integrated emission measure 
$EM_0=EM(t=t_1)$ at the emission measure peak time $t_1$.
This yields the mean electron density in the CME source volume
according to $n_e = \sqrt{EM_0/V}$. For the pre-flare or pre-CME 
phase, which serves to measure the CME mass, we find a very narrow
distribution with a mean of $n_e=(1.2\pm0.5) \times 10^9$ cm$^{-3}$
(Fig.~5c).

From the electron density $n_e$ and the CME source volume $V_0$ we
can then directly calculate the CME mass with 
\begin{equation}
	m_{cme} = m_p n_e V_0 
		\approx 0.81 \times 10^{15} \ 
		\left( { L \over 10^{10}\ {\rm cm}} \right)^2
		\left( { n_e \over 10^{9}\ {\rm cm}^{-3}} \right)
		\left( { T_0 \over 1\ {\rm MK}} \right) \quad [g] \  .
\end{equation}
where we inserted the volume definition $V=L^2 \lambda_{n_e}(T_0)$ (Eq.~(1), 
and $m_p$ is the proton mass. The distribution of CME masses
is shown in form of power law distributions in Fig.~(4e), which
exhibits a total range of $m_{cme}=(0.07-31) \times 10^{15}$ g. 

\subsection{	Fitting of Emission Measure Profile	}

The observed EUV dimming profile $EM(t)$ (e.g., Fig.~3e) exhibits
generally a steep rise before the CME launch at time $t_1$ (defined
by the peak time of the emission measure profile), which
then monotonically drops afterwards, which we interpret as 
density upflow by chromospheric evaporation (during the rise of the
total emission measure), followed by EUV dimming caused by 
adiabatic expansion of the CME volume (during the decay phase).
Our theoretical model of adiabatic expansion (Eqs.26-28) can be
fitted to the observed EUV dimming profile $EM(t)$ with four 
free parameters: the acceleration constant $a_1$,
the acceleration start time $t_1$, 
the acceleration time interval $\tau_{acc}=t_2-t_1$, and
the background fraction level $q_{bg}$. 
We fit these four free parameters
to the EUV dimming profile $EM(t)$ for each event in a time range
$[t_{f1},t_{f2}]$ (marked with red lines in Fig.~3), where the
beginning of the fitting interval coincides with the peak of the
emission measure, $t_{f1}=t_1$, and the end of the fitting interval
is $t_{f2}=t_1+2 \tau_{acc}$, which corresponds to the double duration
of the acceleration time interval $\tau_{acc}$. This time
interval $[t_{f1},t_{f2}]$ covers the steepest decrease of the 
EUV dimming profile in a symmetric way, where the detection of
the dimming is most significant, while the dimming-related
EUV emission drifts outside the field-of-view of the AIA images
later on (at a distance of $\gapprox 1.3 R_{\odot}$), 
a second-order effect that is not modeled here. 

The distributions of the best-fit parameters is shown in Figs.~4 and 5,
including the acceleration constant $a_1$ (Fig.~4g),
the acceleration time interval $\tau_{acc}=$ (Fig.~4h), and
the background fraction level $q_{bg}$ (Fig.~5f). 
the peak fraction level $q_{peak}$ (Fig.~5g), 
and the fit quality $q_{fit}$ (Fig.~5h), which is a measure of the mean
deviation between the observed and modeled EUV emission measure,
normalized by the maximum emission measure $EM_{max}$. The
accuracy of the fits is typically 5\% of the peak emission measure.

\subsection{	CME Acceleration Parameters 			}

The most important fitting parameter is the acceleration time $\tau_{acc}$,
which defines the end time $t_2$ of the acceleration phase and 
the distance $x_2$ (Eq.~14) of the CME at the time of maximum
velocity $v_2=v(t=t_2)$, when the acceleration stops and deceleration due to the
solar gravity sets in. We can compare the best-fit parameters $v_2$
with those estimated from the energy equi-partition theorem,
based on the flare temperature $v_{2T}(T_w)$ (Eq.~30), or based on the 
CME mass $v_{2m}(m_{cme})$ (Eq.~35). The distributions of the velocities
$v_{2T}$ and $v_{2m}$ are shown in Fig.~(5d) and (5e),
covering ranges of $v_{2T}=340-930$ km s$^{-2}$, and  
$v_{2m}=340-1540$ km s$^{-2}$, respectively (Table 1). 

Having the acceleration time interval $\tau_{acc}=t_2-t_1$ and
the maximum velocity $v_2$ established, we obtain immediately
the acceleration rate, $a = v_2/\tau_{acc}$ (Eq.~14) and the 
acceleration height $h_2=r_2-R_{\odot}=r_0-R_{\odot}+(a/2) 
\tau_{acc}$ (Eq.~14), since we assumed constant acceleration 
during the acceleration phase in our model. 
The distributions or the acceleration rate ($a_1=0.24-10$ km s$^{-2}$)
(Fig.~4g), the CME acceleration time ($\tau_{acc}=120-3360$ s),
(Fig.~4h), and the acceleration height $h_2=(0.11-2.38) \times R_{\odot}$
(Fig.~4i), which all show power law-like distributions 
(over a relatively small range of 1.5 decades).

\subsection{	Eruptive and Confined CMEs 		}

The time $t_2$ of maximum CME speed at distance $x_2$ is the
earliest time when it can be decided whether the CME is eruptive
or confined, simply by comparing the velocity $v_2$ with the
local escape speed $v_{esc}(x_2)$ (Eq.~21). The height dependence
of the escape speed due to the $r^2$-dependence of the gravitational
force is shown for one case in Fig.~(3c), which varies from
$v_{esc}(r_0) \approx 550$ km s$^{-1}$ to  
$v_{esc}(r_2) \approx 350$ km s$^{-1}$ at the height $r_2$
of maximum CME speed. The CME has a maximum velocity of 
$v_2=619$ km s$^{-2}$, and thus is eruptive in this case.

In our statistics of 860 CME events, we find 841 eruptive
CME events, and 19 confined flares ($\approx 2.3\%$). 

\subsection{	LASCO and AIA Event Association  	}

For a comparison of AIA results with LASCO data of CMEs we have first
to evaluate which events are associated. The primary time definition
of the analyzed events comes from the GOES flare catalog, which
defines a start, peak, and end time for each of the analyzed flare events.

The LASCO data are catalogued in a CME event list that is available
online, {\sl https://cdaw.gsfc.nasa.gov/ CME$\_$list/},
and their detection time is defined when the
CME leading edge shows up the first time at the edge of the LASCO
occulter disk at a distance of $x_3=4 R_{\odot}$ from Sun center.
The delay of the LASCO detection and the CME launch as detected 
by the EUV dimming in AIA data is expected to vary over a range of
$t_{delay} = 4 R_{\odot} / v_{cme} \approx [0.4, 4]$ hrs, for 
CME speeds in the range of $v_{cme}=[200, 2000]$ km s$^{-1}$.

In Fig.~6a we show a histogram of the expected time delays
$t_{AIA}$ based on our CME model described in Section 2, which
exhibits a mean value of $t_{AIA}=0.39 \pm 0.23$ hrs. An upper
limit above 3 standard deviations would be around 
$t _{AIA} \lapprox 0.39 +3 \times 0.23 =1.1$ hrs. 
On the other side, the histogram of time delays
$t_{LASCO}$ observed with LASCO show a similar almost Gaussian
distribution with a mean of $t_{LASCO}=0.44 \pm 0.25$ hrs (Fig.~6b).
If we fit a Gaussian function, we estimate an upper limit of 
$t_{LASCO} \lapprox 0.44 + 3 \times 0.25 \approx 1.2$ hrs.
Since the CME propagation delay has to be positive by definition,
we expect that the range of physical delays is bound by
the range $t_{AIA} \approx t_{LASCO} \lapprox 1.0$ hrs. 
Most of the AIA events (828) are found in this range,
while LASCO exhibits a smaller number of 432 CME detections inside
this range (52 \%).  A scatterplot of the AIA and LASCO detection 
delays is shown in Fig.~6c, which shows a mean difference of
$t_{AIA}-t_{LASCO} = 0.05 \pm 0.50$ hrs, which implies a slight
systematic error in the inferred CME speeds (either from AIA,
or LASCO, or both).

\subsection{	LASCO Outliers of CME Masses 	  	} 

We measure the CME mass here with two different methods, either with
the conventional method based on the white-light polarized brightness 
in coronagraph images, or with the novel method using the EUV dimming.
We were able to measure a
EUV dimming effect, and therefore a CME mass, in all 860 flares 
from AIA/SDO data, while the white-light data from LASCO/SOHO 
reported a CME detection in about 432 events thereof. 

We show a histogram of the CME masses obtained with LASCO in
Fig.~7a, the corresponding CME masses inferred from AIA in
Fig.~7b, and a scatterplot of the two types of masses in Fig.~7c.
The range of LASCO-inferred CME masses covers a range of
$m_{LASCO}=(0.01-37) \times 10^{15}$ g, and the AIA-inferred CME
masses cover a narrower range of $m_{AIA}=(0.1-30) \times 10^{15}$ g.
Thus the two ranges exhibit a very similar upper limit, but differ
in the lower limit. It appears that LASCO under-estimates CME masses
up to an order of magnitude. The sample of CME masses measured with
AIA exhibits a sharp cutoff around $m_{cme} \approx 0.25 \times 10^{15}$ g
(dashed line in Fig.~7b, while LASCO reports CME masses smaller than
this lower limit for 99 events.
We inquired the LASCO CME catalog and found that most of these
low-mass events were evaluated as ``poor'' or ``very poor'' quality
in the LASCO CME event catalog. These poor events were identified 
near the LASCO detection threshold, which is likely to
be the cause of under-estimating the CME mass. When we apply the
same limit to the AIA-inferred masses, there is only one maverick
event below $m_{AIA}=0.25 \times 10^{15}$ g (see dashed lines in
Fig.~7). So, there are three reasons why these low-mass cases obtained
with LASCO data are likely to be outliers: (1) The fact that AIA detects
no CME masses below the limit of $m_{cme} = 0.25 \times 10^{15}$ g;
(2) AIA should detect lower CME masses because the EUV dimming method 
based on the total emission measure is more sensitive than the
polarized brightness method in white light (because we detect
EUV dimming in all of the joint 456 events, while LASCO detects CMEs
unambiguously in 348 ($\approx 76\%$) events only; and (3) most 
of the low-mass LASCO events suffer from instrumental sensitivity 
problems (qualified as ``poor'' or ''very poor'' events, mostly
detected in the coronagraph C2 only).  In the following analysis 
we will discard these low-mass outliers, which leaves 334 events 
for further LASCO and AIA comparisons. 

We determine the average mass ratio between AIA and LASCO-inferred
CME masses and find $m_{AIA}/$ $m_{LASCO}=0.92$, with a standard 
deviation factor of 3.6 of the logarithmically averaged masses
(Fig.~7d). The lower mass limit apparently yields a closer agreement
between the two instruments, because the ratio is less
commensurate when the low-mass outliers are included.
Applying the lower mass limit, the CME
masses cover a range of $m_{cme}=(0.25-37) \times 10^{15}$ g,
which corresponds to a mass variation by a factor of $\approx 150$. 
The cross-correlation coefficient between the LASCO and AIA-inferred
CME masses is $CCC \approx 0.29$ (Fig.~7d), indicating a weak
correlation due to the remaining methodical errors in modeling of the
CME mass and the measurement of underlying parameters.

\subsection{	Scaling Laws of CME Parameters			}

The CME speed is traditionally measured with the white-light method 
from coronagraphic observations, typically at distances of a few
solar radii when the CME emerges behind the occultation disk,
e.g. at $x_3 \gapprox 4 R_{\odot}$ with LASCO. In contrast, the
EUV dimming effect allows for measurements of the CME expansion
speed in the lower corona, most sensitive at altitudes that
correspond to one emission measure scale height. Extrapolating
the time evolution of CME speeds to heliospheric distances as
probed by coronagraphs, requires additional constraints.

In Section 2.7 we derived a prediction of the maximum speed
based on the energy equi-partition model between the kinetic and 
the thermal flare energy, which predicts a relationship 
\begin{equation}
	v \propto T_e^{1/2} \ .
\end{equation} 
The scatterplot of CME speeds $v_3$ 
(calculated with the relationship of Eq.~24 at the coronagraphic 
occulter position $x_3=4 R_{\odot}$) with the flare emission 
measure-weighted flare temperatures $T_e$ is shown in Figs.~8a
(for LASCO speeds) and in Fig.~9a (for AIA speeds),
which agree within factors of 1.7 and 2.3, respectively. 

As an alternative model we employed the Rosner-Tucker-Vaiana
scaling law (Section 2.8), which predicts a relationship
between the CME mass and the flare temperature, which can be
used to predict a relationship of the CME mass with the
maximum CME speed, i.e., 
\begin{equation}
	v \propto (m_{cme})^{1/4} \ .
\end{equation} 
We show a scatterplot of the CME velocity $v_3$ (calculated 
at location $x_3$ with Eq.~24) with the CME mass (raised 
to the power of 1/4) in Figs.~8b (for LASCO speeds) and Fig.~9b 
(for AIA speeds). We find a good agreement within
a factor of 2.7 and 2.6, respectively.
The cross-correlation coefficient is $CCC=0.62$ (Fig.~8b),
which confirms that the RTV relationship indeed is physically 
related to the CME speed. 

The correlations of LASCO and AIA CME speeds $v_2$ or masses $m_{cme}$ 
with the CME length scales $L$ are shown in Figs.~8c,d and Fig.~9c,d.
The tightest correlation is found between the AIA-inferred
CME mass and the CME length scale $L$, with a cross-correlation
coefficient of $CCC=0.77$ (Fig.~9d), which reflects the
scaling law of the RTV relationship between mass and length
scale,
\begin{equation}
	m_{cme} \propto L^2 \ .
\end{equation}
This scaling law follows from the definition of the CME mass,
$m_{cme} \propto A^2 h_0$ (Eq.~1), if the density scale height
$h_0$ is constant or has only little variation.

These results allow us to use the RTV-based model as a 
prediction of the maximum CME speed $v_2$ for AIA data, 
solely based on the measurement of the CME mass 
from the EUV dimming, i.e., $v_{AIA} = 
767$ km s$^{-1}$ $\times (m_{cme}/10^{15}$ g)$^{1/4}$.
Based on this formula, a velocity range of $v_2=540-1900$
km s$^{-1}$ is predicted by the RTV law.

\subsection{	Comparison of Old and New CME Model  			}

A scatterplot of the CME parameters obtained with both instruments
AIA and LASCO, which includes the CME speed ($v_{LASCO}$ vs. $v_{AIA}$),
the CME mass ($m_{LASCO}$ vs. $m_{AIA}$), and the kinetic energy
($E_{LASCO}$ vs $E_{AIA}$), has been shown in Fig.~18 for 218 CME
events in the previous Paper IV.
We show the scatterplots of the same parameters obtained with the
new method (Fig.~10, left panels), sampled after elimination
of low-mass and mis-associated events, using the 
equi-partition assumption and the RTV scaling laws.
The substantial scatter between the old and new method
indicates significantly differences for individual events,
but the resulting size distributions are similar
(Fig.~10, right panels).

\section{		DISCUSSION 				}

\subsection{	Measurements of Coronagraph-Detected CMEs 	}

CMEs have traditionally been observed and measured with coronagraphs
in white-light, such as with the {\sl Orbiting Solar Observatory OSO-7} 
(Tousey 1973), the {\sl Apollo Telescope Mount (ATM)} onboard
Skylab (Mac Queen et al.~1974), the Solwind coronagraph onboard
P78-1 (Michels et al.~1980), the {\sl Coronagraph/Polarimeter (CP)}
onboard the {\sl Solar Maximum Mission (SMM)} (House et al.~1980),
the {\sl Large-Angle and Spectrometric Coronagraph (LASCO)}
(Brueckner et al.~1995), the {\sl Solar Mass Ejection Imager
(SMEI)} (Eyles et al.~2003), and the {\sl Sun Earth Connection
Coronal and Heliospheric Investigation (SECCHI)} onboard the
{\sl Solar Terrestrial Relations Observatory (STEREO)} with
the two coronagraphs COR1 and COR2 (Howard et al.~2008). 
What could be measured with these coronagraphs is primarily 
the detection time $t_{cme}$ when the CME emerges from behind 
the coronagraph occulter disk, for instance at a distance of 
$x_{cme} \approx 4 R_{\odot}$ for LASCO, 
their angular width $w_{cme}$, their mass $m_{cme}$ based on 
the polarized brightness produced by Thomson scattering,
and their projected speed measured in the field-of-view 
of the coronagraphs. Once the mass $m_{cme}$ and the speed $v_{cme}$
is known, their kinetic energy $E_{kin}=(1/2) m_{cme} (v_{cme})^2$
can directly be obtained. 

Catalogs of CME events observed with LASCO have been published
and are available on websites, such as the LASCO/CDAW catalog
{\sl https://cdaw.gsfc.nasa.gov/CME$\_$list/}, the LASCO-based
{\sl Computer Aided CME Tracking (CACTUS)} catalog 
{\sl http://sidc.oma.be/cactus} (Robbrecht and Berghmans 2004; 
Robbrecht et al.~2009), the LASCO-based {\sl Solar Eruptive
Event Detection System (SEEDS)} catalog (Olmedo et al.~2008),
and the LASCO-based {\sl Coronal Image Processing (CORIMP)} 
catalog (Byrne et al.~2012; Morgan et al.~2012). While these
catalogs are all based on LASCO-detected events, there are also
CME measurements using the STEREO/COR2 instrument given in
the SEEDS and CACTUS catalogs. 

What are the strengths and weaknesses of the coronagraph-based
measurements of CMEs? The foremost advantages of coronagraphic
measurements are: (1) The uninterrupted long-term availability 
over 20 years
(since the launch of SOHO in 1996); (2) the derivation of the
CME mass is independent of the temperature; (3) the measurement
of the projected leading-edge CME speed is well-defined and 
can be automated (for instance with the Hough transform;
Robbrecht and Berghmans 2004). On the other side, there are a
number of disadvantages that makes the EUV dimming method truly
complementary: (4) The association of CMEs with flare events is
often ambiguous; (5) The kinematics or the evolution of the
height $h(t)$, the velocity $v(t)$, and acceleration $a(t)$
in the source region is not known in the altitude range where
a CME is occulted by the coronagraph disk 
(at $x_3 \lapprox 1.6-4.0 R_{\odot}$); and (6) the magnetic
topology and plasma temperature diagnostics in the CME
source region cannot be deduced from white-light data;  
(7) Projection effects in halo CMEs make it difficult to
disentangle the kinematics and 3D geometry of Earth-directed CMEs.
All these deficiencies make it also difficult to derive
3D models of CMEs that cover the entire evolution from the
source region in the lower corona out to the heliosphere,
which is a necessary pre-requisite to test data-driven
MHD simulations and theoretical models of CMEs. 

\subsection{	Measurements of CMEs from EUV Dimming 		}

Measurements of CME parameters from EUV dimming information 
started with the availability of solar EUV images, such as
with the {\sl EUV Imaging Telescope (EIT)} onboard SOHO since 1996 
(e.g., Thompson et al.~2000), the {\sl Extreme Ultra-Violet Imager
(EUVI)} onboard STEREO since 2006 (e.g., Bein et al.~2011), 
and the {\sl Atmospheric Imager Assembly (AIA)} onboard SDO 
since 2010 (e.g., Cheng et al.~2012; Mason et al.~2014;
Kraaikamp and Verbeek 2015; Aschwanden 2016, Paper IV). 
Dimming regions were identified by areas of strong depletion 
in the EUV brightness, mapping out the apparent ``footpoint'' area 
of a CME, which is detected with a white-light coronagraph 
generally about an hour later (Thompson et al.~2000). Consequently,
a high association rate of $\approx 55-74\%$  was found between
EUV dimming and CME events (Bewsher et al.~2008; Nitta et al.~2014).
The 3D structure of CME source regions and the associated EUV 
dimming could be modeled with stereoscopic methods 
(Aschwanden et al.~2009a,b; Aschwanden 2009; Temmer et al.~2009; 
Bein et al.~2013). A key result was that the CME mass determined 
from EUV dimming agreed well with those determined with the 
white-light scattering method ($m_{\mathrm{EUVI}}/
m_{\mathrm{LASCO}}=1.1 \pm 0.3$), and agreed also between the two
STEREO spacecraft A and B ($m_A/m_B=1.3 \pm 0.6$) (Aschwanden
et al.~2009a). Another benefit of stereoscopic observations is the
determination of the 3D trajectory and de-projected CME speed and mass
(Bein et al.~2013). 

Systematic measurements of the main physical parameters of
EUV dimming events and associated CMEs started only recently,
amounting to a statistical study of 399 events during the first
3.5 years of the SDO mission (2010-2014; Paper IV), which we 
expand here to 864 events during the first 7 years of the SDO
mission. The basic measurements consist of the decreasing slope
in the total emission measure profile of the EUV brightness $EM(t)$ 
during a flare, which can be modeled with a spatially expanding
CME volume $V(t)$ and the assumption of an adiabatic process. 
In the simplest scenario, the time evolution of the CME volume
is reciprocal to the mean electron density, $n_e(t) \propto 
V(t)^{-1}$, which can be related to the volume-integrated
emission measure by $EM(t) \propto n_e(t)^2 V(t)$ for optically
thin plasmas. In the previous study (Paper IV), the velocity of
the expanding CME volume was derived from forward-fitting of
the systematically decreasing emission measure profile $EM(t)$
after the flare peak, which turned out to be strongly disturbed 
by multiple brightenings immediately followed by dimming episodes 
in large and complex flare events. Hence, we developed a new
method in this study where the CME velocity is determined
from the equi-partition between kinetic CME velocity and
the thermal flare energy, which predicts a simple relationship
between the CME velocity and the flare temperature, $v_{cme} 
\propto (T_w)^{1/2}$. The emission measure-weighted flare
temperature can directly be obtained from the DEM distribution
of the flare region obtained with an automated algorithm in
the flare/dimming region (Aschwanden et al.~2013).
Moreover, we derived a redundant
method where the RTV scaling law is applied to the flare
loops at the peak time (turnover point) when the heating 
rate and the cooling rate is balanced, which yields a
simple relationship between the CME velocity and the CME
mass, $v_{cme} \propto (m_{cme})^{1/4}$. With these new
developments we obtained a CME model that is completely
based on EUV dimming data and can predict all CME white-light
parameters as well as additional CME model parameters with
unprecedented robustness. 

\subsection{	CME Acceleration and Deceleration 	} 

Our CME model provides a powerful tool to diagnose acceleration
and deceleration phases of CMEs. Acceleration can be caused by
the magnetic pressure term of the Lorentz force, a pressure
gradient, and the solar wind flow, while deceleration can be caused 
by the Sun's gravity, the aerodynamic drag, and the tension of 
the magnetic field, as studied by 3D MHD simulations 
(e.g., Shen et al.~2012). 

There is an increasing number of observational studies available
now that provide statistical information on CME acceleration.
The range of of acceleration rates is quite different near
the Sun, typically $a_{max} \approx 100-2000$ m s$^{-2}$
(e.g., Zhang and Dere 2006; Cheng et al.~2010; Bein et al.~2011; 
Joshi and Srivastava 2011),
compared with the heliosphere, say in the LASCO field-of-view at
$r \gapprox 4 R_{\odot}$, where it can be positive or negative,
typically in a range of $a=\pm20$ m s$^{-2}$ 
(e.g., Michalek 2012). For the solar gravity force alone
we would expect a deceleration of $a=-\Gamma M_{\odot}
/(4 R_{\odot})^2 \approx -17$ m s$^{-1}$ at the inner 
boundary of the LASCO field-of-view. 

How does our EUV dimming model complement previous measurements
of CME acceleration and deceleration? Our AIA-constrained dimming 
model can fill in the gaps of measurements below the coronagraph
occultation height (say at $r \lapprox 4R_{\odot}$ for LASCO),
which contains the most important height range for studying the
magnitude and duration of magnetic Lorentz forces that accelerate
the CME. Secondly, the AIA-constrained method yields the total
emission measure, which contains all particle contributions in
the temperature range of $T_e \approx 0.5-20$ MK, so that the
particle number is almost completely conserved, similar to the
white-light Thomson scattering method, and thus heating or cooling
processes do not interfere with the detection of EUV dimming. 
The broad AIA temperature
coverage also constrains the emission-measure weighted temperature
in the flare region, which is found here to be a good predictor of the
maximum CME speed (Eq.~30). Furthermore, our analytical EUV-dimming 
model allows us to measure the duration and magnitude of the
acceleration rate, as well as the gravity-driven deceleration,
which are the most dominant force components for weak CMEs 
propagating near the escape speed.  

\subsection{	Confined and Eruptive CMEs 		} 

Since both acceleration and gravity-driven deceleration is
built in our analytical model, we should be in a good position
to discriminate between eruptive and confined CMEs. For our
analyzed data set of 860 flare events we found only 19 events 
(2.3\%) to be confined flares. This is a relatively low
percentage, compared with other studies. For instance,
Cheng et al.~(~010) analyzed a sample of 1246 flare events
and found that 706 events (57\%) are associated with CMEs,
while the other 540 flares (43\%) are confined. The
discrepancy may be related to the detection method. 
Cheng et al.~(2010) did a visual inspection of the movies
observed by LASCO and EIT/SOHO and identified CME-associated 
flare events if the spatio-temporal co-registration of
transient flare brightenings and large-scale dimming
on EIT images occurs. That means that confined flares
are defined by the absence of spatio-temporal coincidences.
It is not clear why this method produces a 25 times higher 
fraction of confined flares than our method of calculating
the maximum CME speed and comparing it with the co-spatial
escape speed. Nevertheless, such discrepancies provide
important tests to sort out methodical biases in the data
analysis and modeling of CMEs. 

\section{		CONCLUSIONS 			}

In this study we refined the method of calculating physical
parameters of CMEs based on the EUV-dimming method, which 
provides a complementary approach to the traditional
white-light method. An extensive study that compares the
two strategies of white-light scattering and EUV dimming
in the measurement of CME parameters 
has been undertaken in the previous Paper IV of this
series on the global energetics of solar flares and CMEs.
Here we focus on the improvements between the previous
EUV dimming method (Paper IV) and the refined method
presented in this study. The methodical improvements and 
related conclusions are summarized in the following.

\begin{enumerate}

\item{{\bf Larger Statistics:} By extending our analysis
from the first 3.5 years of the SDO mission (with 399 events)
to the currently entire SDO era of 7 years (2010-2016) 
(with 864 events) doubles the size of the statistical data 
sample (containing M and X-class flare events) investigated 
here, for both the AIA/SDO and LASCO/SOHO data sets. Most
previous studies apply CME models to small samples of 
observed CMEs only and are therefore not statistically
representative.}

\item{{\bf The Spatial CME Geometry:} is given by self-similar
(adiabatic) 3-D expansion of a spherical volume in the refined 
model, while the old model assumed the self-similar 1-D expansion 
of a wedge. The 3D spherical geometry appears to be more realistic, 
based on the observation of bubble-like CME geometries on one
side, and the plausibility of isotropic expansion in coronal 
regions with a low plasma-beta parameter on the other side.}

\item{{\bf The Gravity Force:} causes a deceleration of the
expanding CME, which has been neglected in the old model (since
it is not important for fast and large CMEs with speeds in
excess of the escape speed. Inclusion of the gravitational
force during the acceleration of CMEs, in contrast, is important
for small CMEs (associated with M-class flares or lower) and
can reproduce the dynamical behavior of ``failed'' CMEs
properly, which allows the discrimination between eruptive and 
confined CMEs. We find that a fraction of 2.3\% of CMEs 
(of $\ge$M1.0 GOES class) is associated with confined flares.}

\item{{\bf Speed Comparisons of CMEs:} between traditional
white-light observations (where the speed is specified at the
leading edge) and EUV-dimming observations (where the speed is
measured from the center-of-mass motion) need to be
self-consistently modeled. The spherical self-similar expansion model
implies a factor of 2 difference in the speeds of the center-of-mass
motion and the leading-edge motion. The resulting corrections 
amount to a factor of 4 in the kinetic energies (since
$E_{kin} \propto v^2$).}

\item{{\bf The association of LASCO and AIA CMEs:} can only be
properly determined if the time difference between the flare 
onset in the lower corona (coincident with the lauch of a CME)
and the first detection in white-light outside the occultation 
disk of a coronagraph is measured and kinematically modeled.
We find that the typical delay for LASCO observations (beyond
an occultation disk with a radius of $r_{occ} \gapprox 4 r_{\sun}$) 
is $\Delta t \approx 1.0$ hr.}

\item{{\bf LASCO is less sensitive than AIA} in detecting  
small CME events. We estimate that LASCO detects $\approx 50\%$
of the AIA EUV dimming events (of $\ge M1.0$ GOES class).}

\item{{\bf The equi-partition} between the CME kinetic energy 
and the thermal flare energy yields a simple scaling law
between the (emission measure-weighted) flare temperature
and the CME speed, i.e., $v_{2T} \propto T_w^{1/2}$, which
provides a robust estimate for extrapolated CME speeds at
heliospheric distances (within a factor of $\approx 2$).
The equivalence of CME energies and thermal energies has
also been established in a previous study (Paper V), where 
the CMEs were found to dissipate $E_{cme}/E_{mag}=0.07\pm0.14$
of the magnetic energy (in the statistical mean of 157 events),
while the thermal energy owns a ratio of $E_{th}/E_{mag}
=0.08\pm0.13$ in 170 events (Table 3 in Paper V), which implies an
equi-partition of $E_{cme}/E_{th}=1.0\pm0.2$ between the
CME kinetic and thermal energy.}

\item{{\bf The Rosner-Tucker-Vaiana (RTV) law}, which is based
on the equi-partition of heating and cooling rates at the flare
peak times yields two simple scaling laws, one between CME mass and
the CME velocity, i.e., $v \propto m_{cme}^{1/4}$, and one between
CME masses and CME footpoint area $A=L^2$, i.e., $m_{cme} \propto L^2$.
Both scaling laws can be used to provide estimates of
CME parameters (within a factor of $\approx 2$).}

\item{{\bf LASCO is under-estimating CME masses:} in 24\% 
of CME events associated with $\ge$M1.0 class flares. From AIA
measurements we estimate that LASCO-inferred CME masses below
a limit of $m_{cme} \lapprox 10^{14}$ g represent under-estimates.}

\end{enumerate}

In summary, the chief advantage of the EUV-dimming method described
and refined in this study is the independent corroboration of the
traditional white-light method to quantify
basic physical parameters of CMEs. Since both methods have unknown
systematic errors, the statistical comparison of the two independent
methods can elucidate and quantify model uncertainties and systematic
errors. The fact that both the white-light and the EUV dimming model
agree in the determination of CME masses and speeds
(within a factor of $\approx 3$) gives us confidence on the 
statistical consistency of both methods, which helps us also to
identify outliers or the parameter space where the models break down.
For instance, CME masses below a limit of $m_{cme} \lapprox 10^{14}$ 
g appear to be systematically under-estimated with the white-light
method. Moreover, the EUV dimming method appears to be more sensitive
than the white-light method for small events ($\gapprox$M1 GOES class). 
The superior sensitivity of CME detection using EUV-dimming data
enables us to measure CME parameters with AIA/SDO in many cases where
the white-light method using LASCO/SOHO data is not available or
is affected by ambiguous timing in the flare association. 
Future case studies of individual events with inconsistent CME parameters
(obtained with either the white-light or the EUV-dimming method) may give
us further insights where present CME models can be improved.

\bigskip
\acknowledgements
We acknowledge useful comments from an anonymous referee
and discussions with 
Nat Gopalswamy, Nariaki Nitta, Manuela Temmer, Barbara Thompson,
Astrid Veronig, Angelos Vourlidas, and Jie Zhang.
This work was partially supported by NASA contract NNG 04EA00C 
of the SDO/AIA instrument. 

\clearpage


\clearpage


\begin{deluxetable}{llllll}
\tablecaption{Ranges and distributions of CME parameters
measured in 860 CME events with AIA. The mean and standard deviations refer to the
slopes $p$ of power law distributions, or to the Gaussian normal distributions
$x\pm \sigma_x$.}
\tablewidth{0pt}
\tablehead{
\colhead{Parameter}&
\colhead{Range}&
\colhead{Median}&
\colhead{Physical}&
\colhead{Distribution}&
\colhead{Mean and}\\
\colhead{}&
\colhead{}&
\colhead{}&
\colhead{Units}&
\colhead{type}&
\colhead{standard}\\
\colhead{}&
\colhead{}&
\colhead{}&
\colhead{}&
\colhead{}&
\colhead{deviation}}
\startdata
Length scale $L$               & $(18, ..., 361)\times 10^8$ 	     & $99 \times 10^8$	    & cm          & power law & $p=-3.7\pm1.2$              \\
CME dimming area $A$           & $(0.03, ..., 13)\times 10^{16}$     & $1.0 \times 10^{16}$ & cm$^2$      & power law & $p=-2.9\pm0.8$              \\
CME dimming volume $V$         & $(0.0005, ..., 0.5)\times 10^{30}$  & $0.02 \times 10^{30}$ & cm$^3$     & power law & $p=-2.8\pm0.7$              \\
CME emission measure $EM/A$    & $(13, ..., 3342)\times 10^{26}$     & $84 \times 10^{26}$  & cm$^{-5}$   & power law & $p=-2.6\pm0.6$              \\
CME mass $m$                   & $(0.07, ..., 31)\times 10^{15}$     & $1.54 \times 10^{15}$& erg         & power law & $p=-2.5\pm0.6$              \\
CME energy $E$                 & $(0.076, ..., 227)\times 10^{30}$   & $3.66 \times 10^{15}$& erg         & power law & $p=-1.6\pm0.2$              \\
Acceleration rate $a_1$        & $(0.24, ..., 10)$                   & 1.43                 & km s$^{-2}$ & power law & $p=-1.7\pm0.2$              \\
Acceleration time $\tau_{acc}$ & $(120, ..., 3360)$                  & 480                  & sec         & power law & $p=-2.6\pm0.6$              \\
Acceleration height $h_2$      & $(0.11, ..., 2.38)$                 & 0.56                 & $R_{\odot}$ & power law & $p=-6.7\pm2.5$              \\ 
                               &                                     &                      &             &                                         \\
Preflare temperature $T_0$ & $(1.2, ..., 5.0)$                   & 1.8                  & MK          & Gaussian  & $T_0=1.8\pm0.3$         \\
Flare peak temperature $T_w$   & $(2.3, ..., 17.4)$                  & 8.9                  & MK          & Gaussian  & $T_w=9.3\pm2.4$             \\
EM scale height $h_0$          & $(0.08, ..., 0.59)$                 & 0.30                 & $R_{\odot}$ & Gaussian  & $h_0=0.31\pm0.08$           \\
Electron density $n_e$         & $(0.28, ..., 6.8)\times 10^9$       & $1.1 \times 10^9$    & cm$^{-3}$   & Gaussian  & $n_e=(1.2\pm0.5)\times 10^9$\\
CME max. speed $v_{2T}$        & $(340, ..., 930)$                   & 665                  & km s$^{-1}$ & Gaussian  & $v_{2T}=670\pm90$           \\
CME max. speed $v_{2m}$        & $(336, ..., 1542)$                  & 766                  & km s$^{-1}$ & Gaussian  & $v_{2m}=590\pm66$           \\
Background fraction $q_{bg}$   & $(0.00, ..., 0.68)$                 & 0.33                 &             & Gaussian  & $q_{bg}=0.22\pm0.12$        \\
Peak fraction $q_{peak}$       & $(0.69, ..., 1.0)$                  & 0.92                 &             & Gaussian  & $q_{peak}=0.89\pm0.08$      \\
Goodness-of-fit $q_{fit}$      & $(0.00, ..., 0.24)$                 & 0.05                 &             & Gaussian  & $q_{fit}=0.07\pm0.15$       \\
\enddata
\end{deluxetable}
\clearpage


\begin{figure}
\plotone{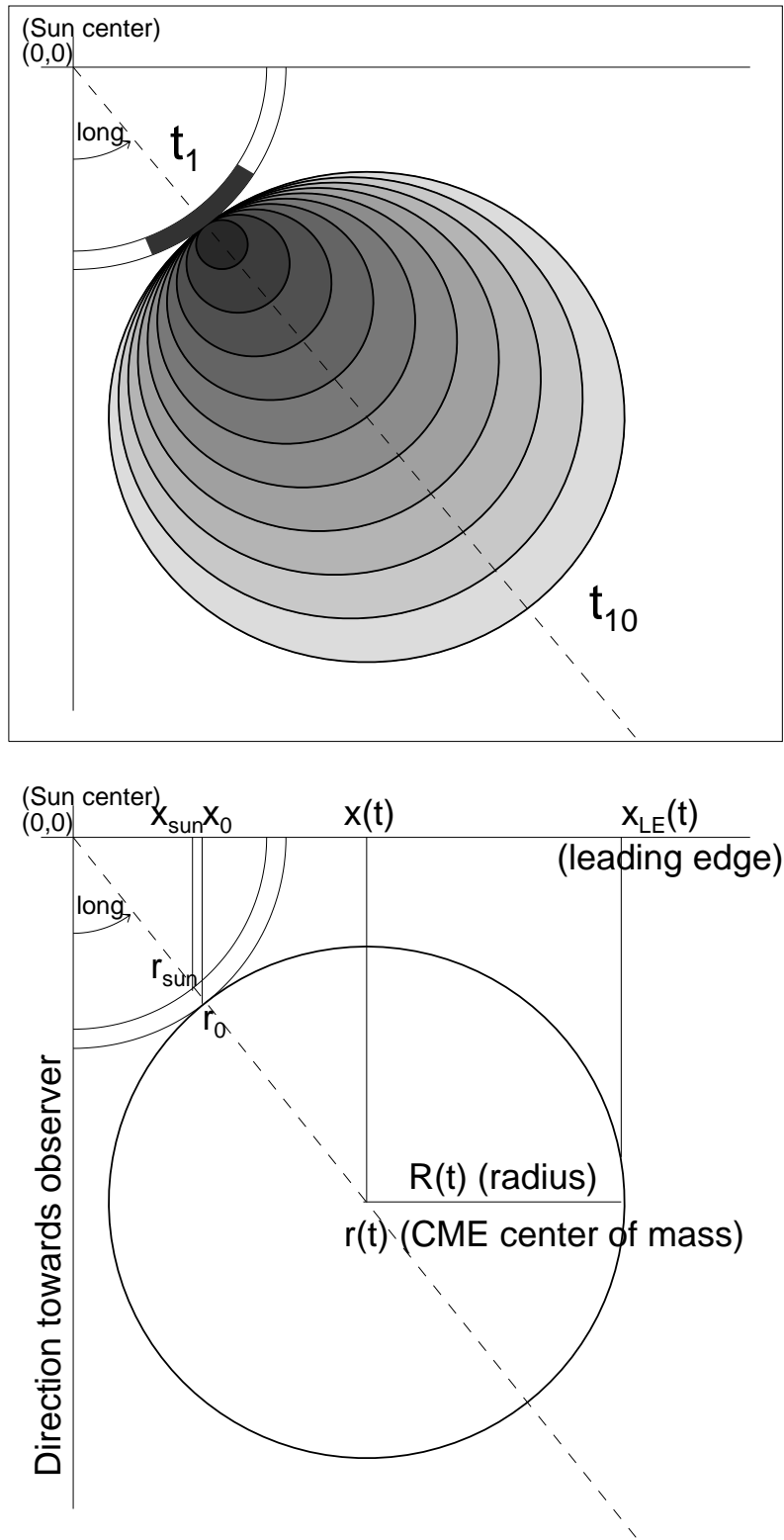}
\caption{Geometric diagram of a spherically expanding CME (top) and
definition of geometric parameters (bottom), along the plane-of-sky (x-axis)
and the radial propagation direction (diagonal dashed line):
$r_{sun}$ and $x_{sun}$ refer to photospheric (and projected) 
distance from Sun center,
$r_0$ and $x_0$ to an altitude of one emission measure scale height
in the corona, $r(t)$ and $x(t)$ to the center of mass of the CME,
$x_{LE}(t)$ to the leading edge of the CME, $R(t)$ is the radius of 
the expanding CME, and $long$ is the longitudinal angle between the 
observer's line-of-sight and the CME center-of-mass trajectory.} 
\end{figure} 

\begin{figure}
\plotone{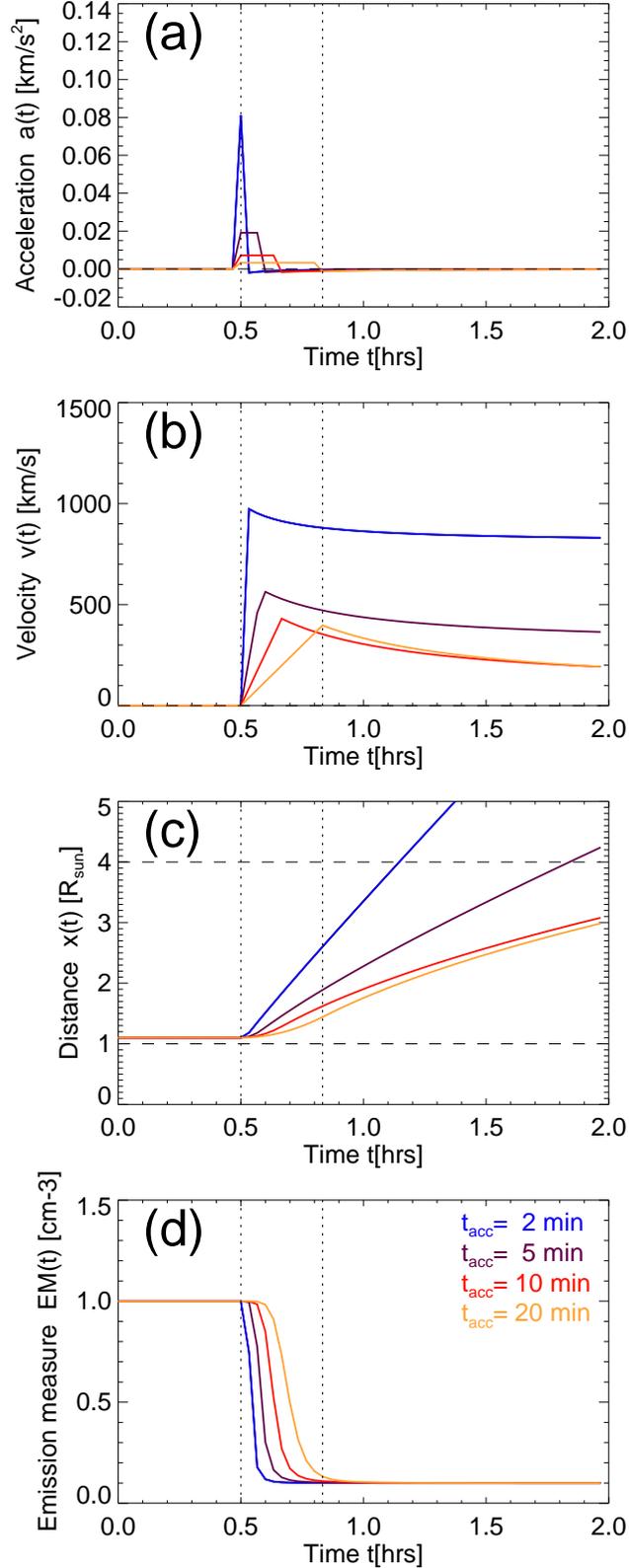}
\caption{Time evolution of: (a) the acceleration $a(t)$, 
(b) the velocity $v(t)$, (c) the projected distance of the CME
center-of-mass motion $x(t)$ (third panel), and (d) the total emission 
measure $EM(t)$, for four different acceleration time intervals
($\tau_{acc}=2, 5, 10, 20$ min). The start $t_1$ and end time $t_2$
of acceleration is indicated with vertical dotted lines for the
case of $\tau_{acc}=20$ min.}
\end{figure}

\begin{figure}
\plotone{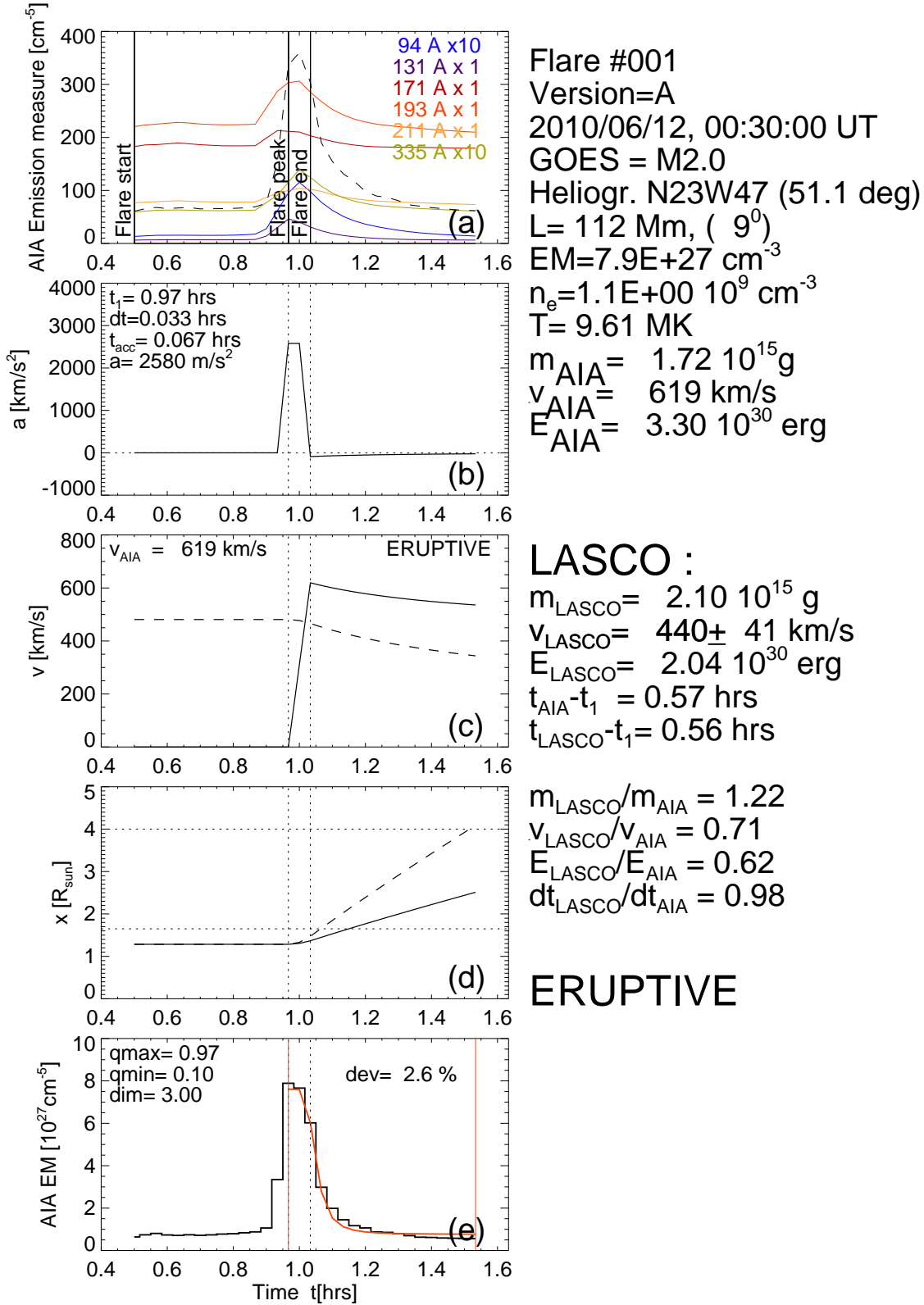}
\caption{Example of data analysis (event \#1), a GOES M2.0-class flare
observed with AIA on 2010 June 12, 00:30 UT: (a) AIA light curves 
(colored) and total emission measure (dashed); (b) acceleration time profile,
(c) CME speed (solid profile) and escape speed (dashed profile);
(d) projected distance of CME center-of-mass (solid curve) and
leading edge (dashed curve), with the initial CME height 
and LASCO coronagraph occultation height
(dotted horizontal lines), and (e) total emission measure
profile $EM(t)$ observed (black histogram) and modeled by forward-fitting
(red thick curve), the fitting time interval (red lines), and
end time of acceleration (dotted).}
\end{figure}

\begin{figure}
\plotone{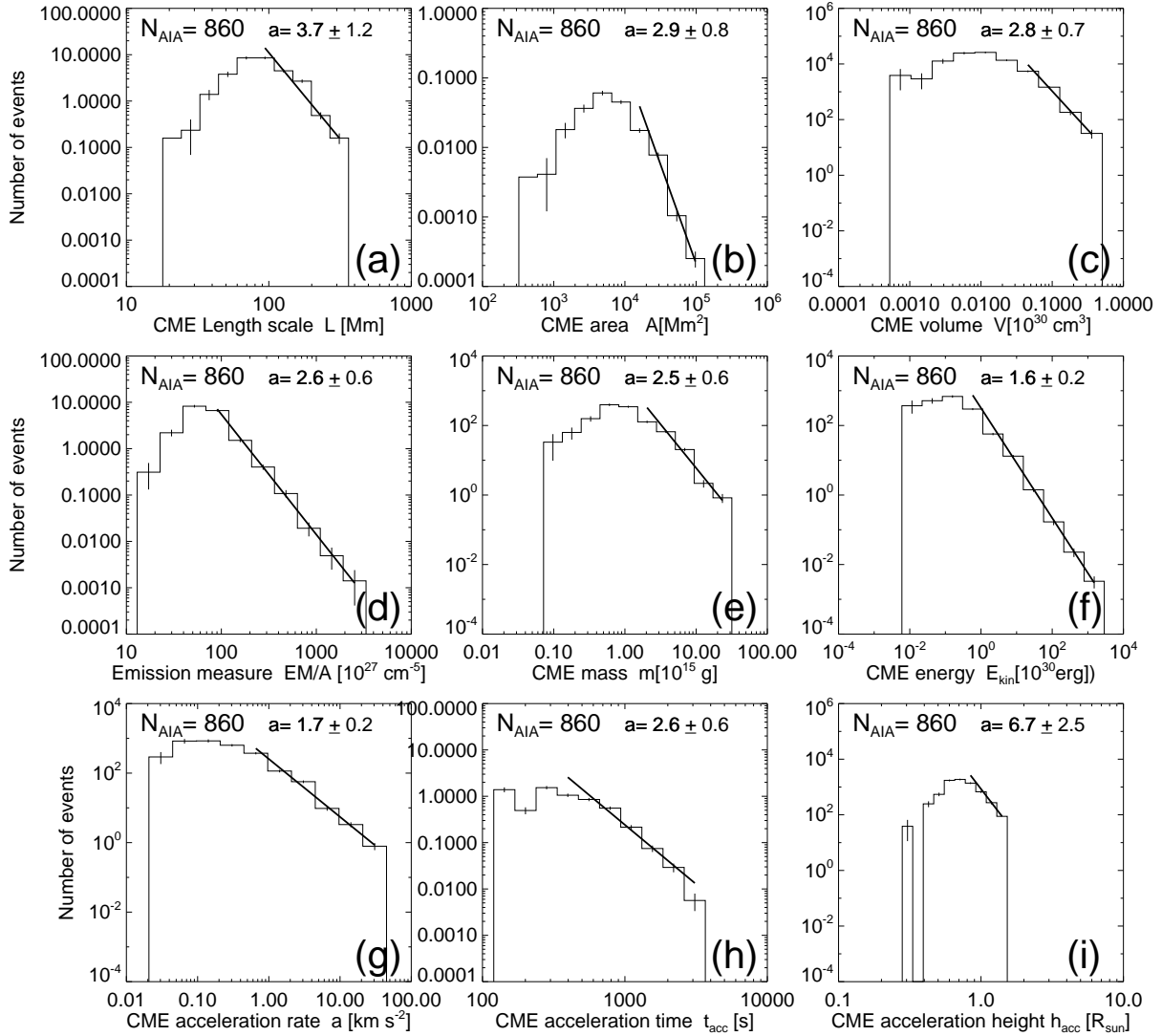}
\caption{Power law-like size distributions of physical parameters in CMEs.}
\end{figure}

\begin{figure}
\plotone{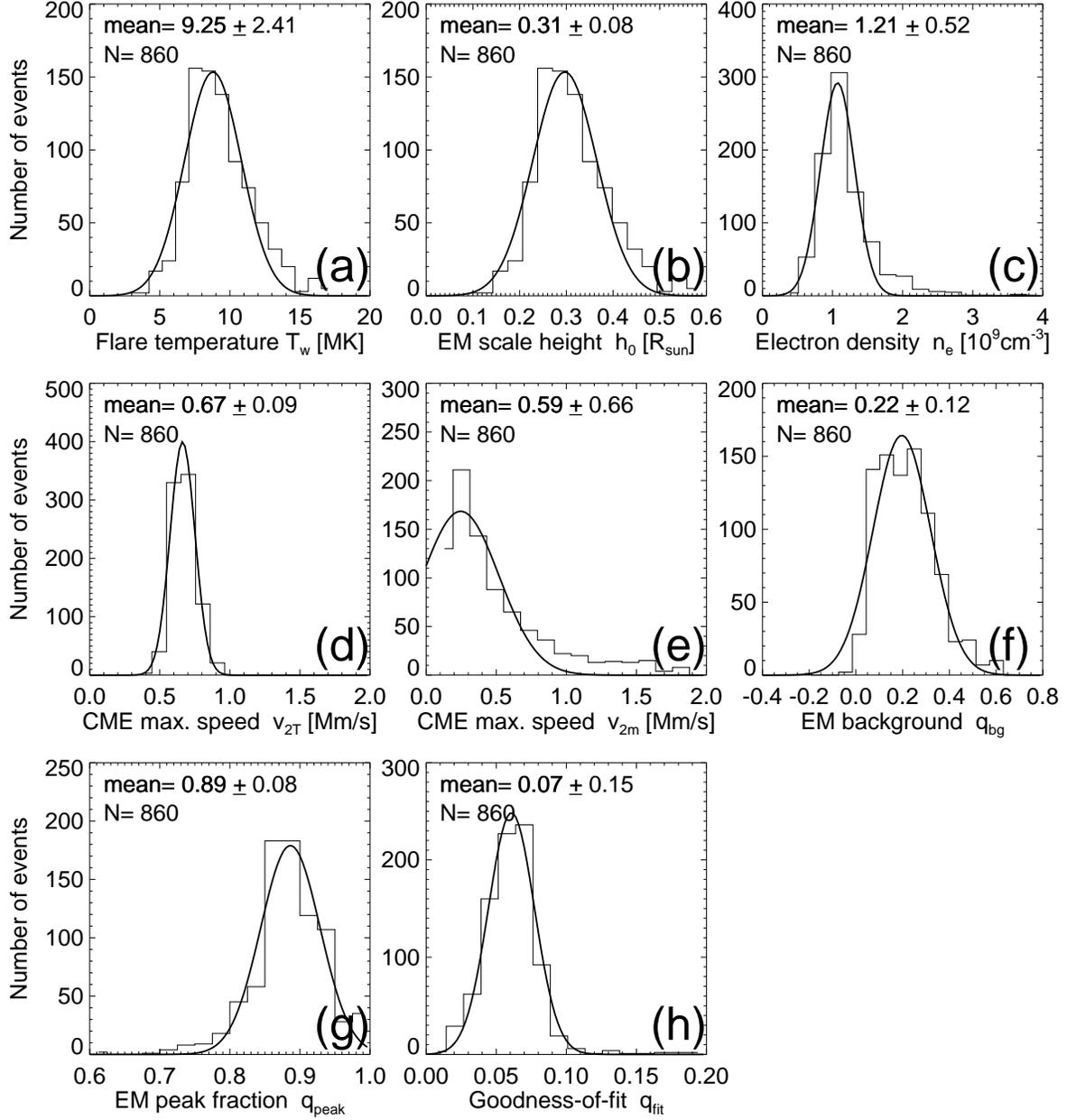}
\caption{Histogram of physical CME parameters and numerical fitting 
parameters with Gaussian-like distributions.}
\end{figure}

\begin{figure}
\plotone{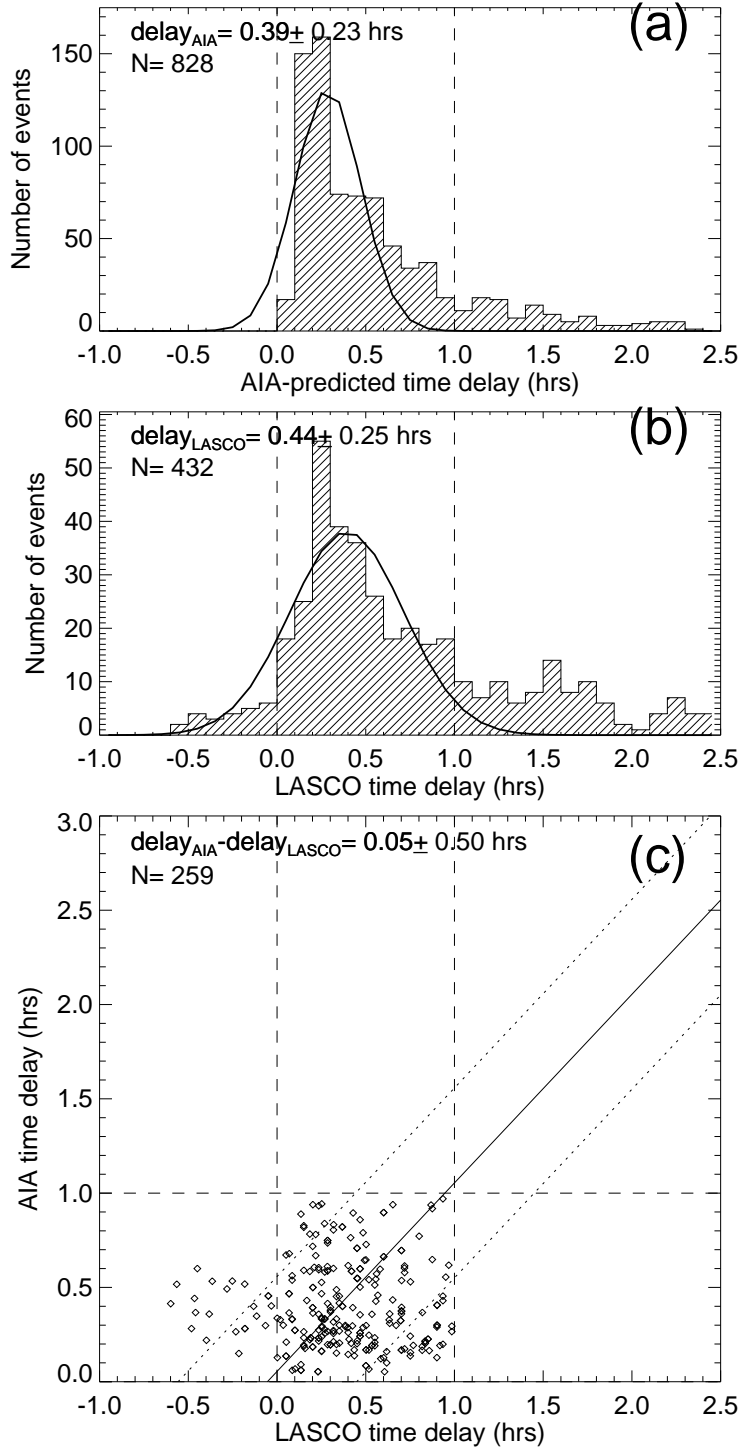}
\caption{Histograms of the CME coronagraphic detection delay
between the start of EUV dimming detected in AIA and the detection
with LASCO at a distance of $x_3=4 R_{\odot}$ from Sun center:
(a) predicted from the AIA-inferred CME model;
(b) observed with LASCO, and
(c) cross-correlation plot. 
A gaussian function is fitted to the histograms.}
\end{figure}

\begin{figure}
\plotone{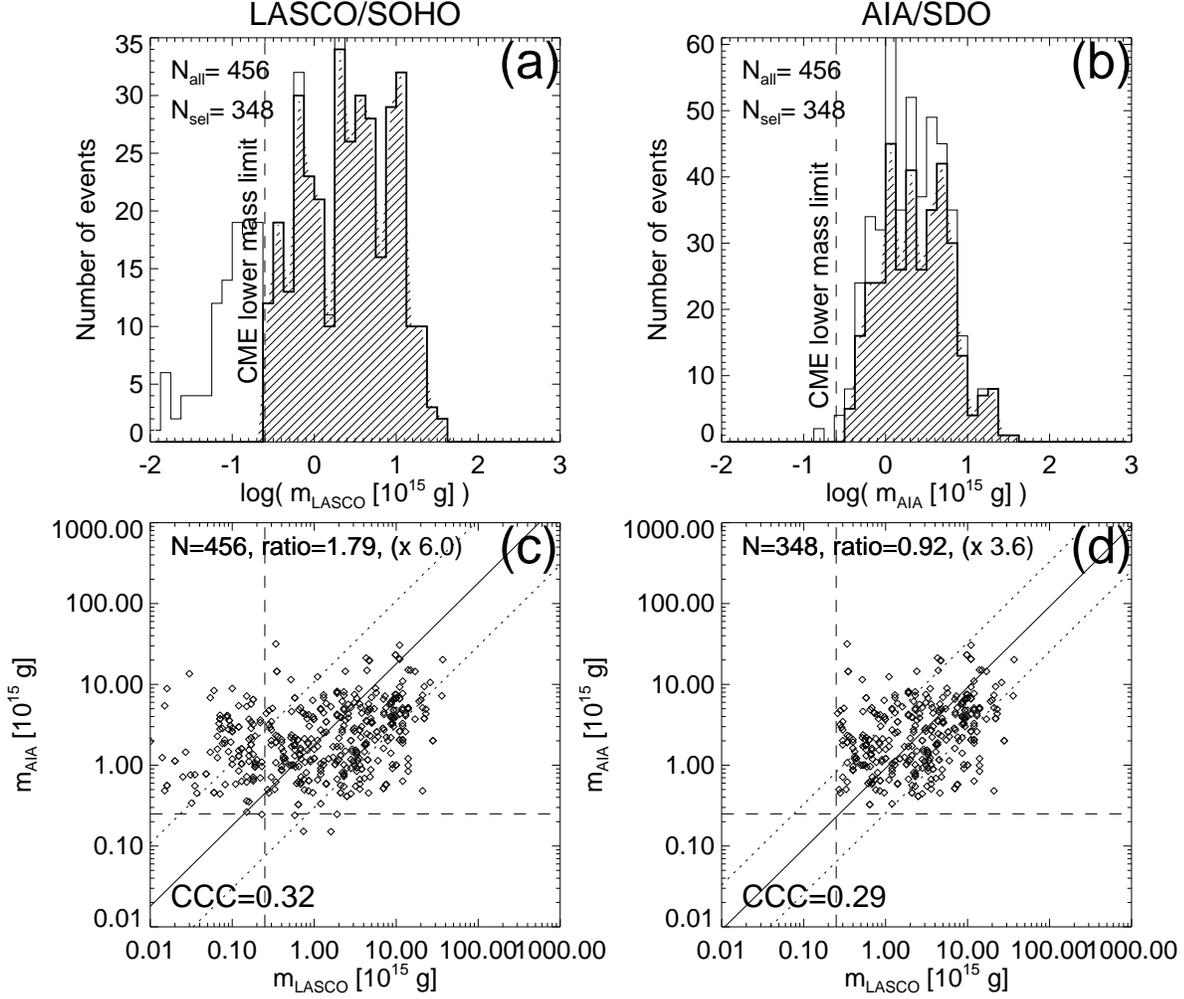}
\caption{(a) Logarithmic histograms of CME masses from LASCO and 
(b) from AIA, (c) scatter plot of AIA-inferred versus LASCO-inferred 
CME masses, (d) with the LASCO under-estimated cases removed. 
The CME lower mass limit at $m \ge 0.3 \times 10^{15}$
is indicated (vertical dashed lines), as well as logarithmic
average ratios (diagonal solid lines) with one standard deviation
factors (diagonal dotted lines).} 
\end{figure}

\begin{figure}
\plotone{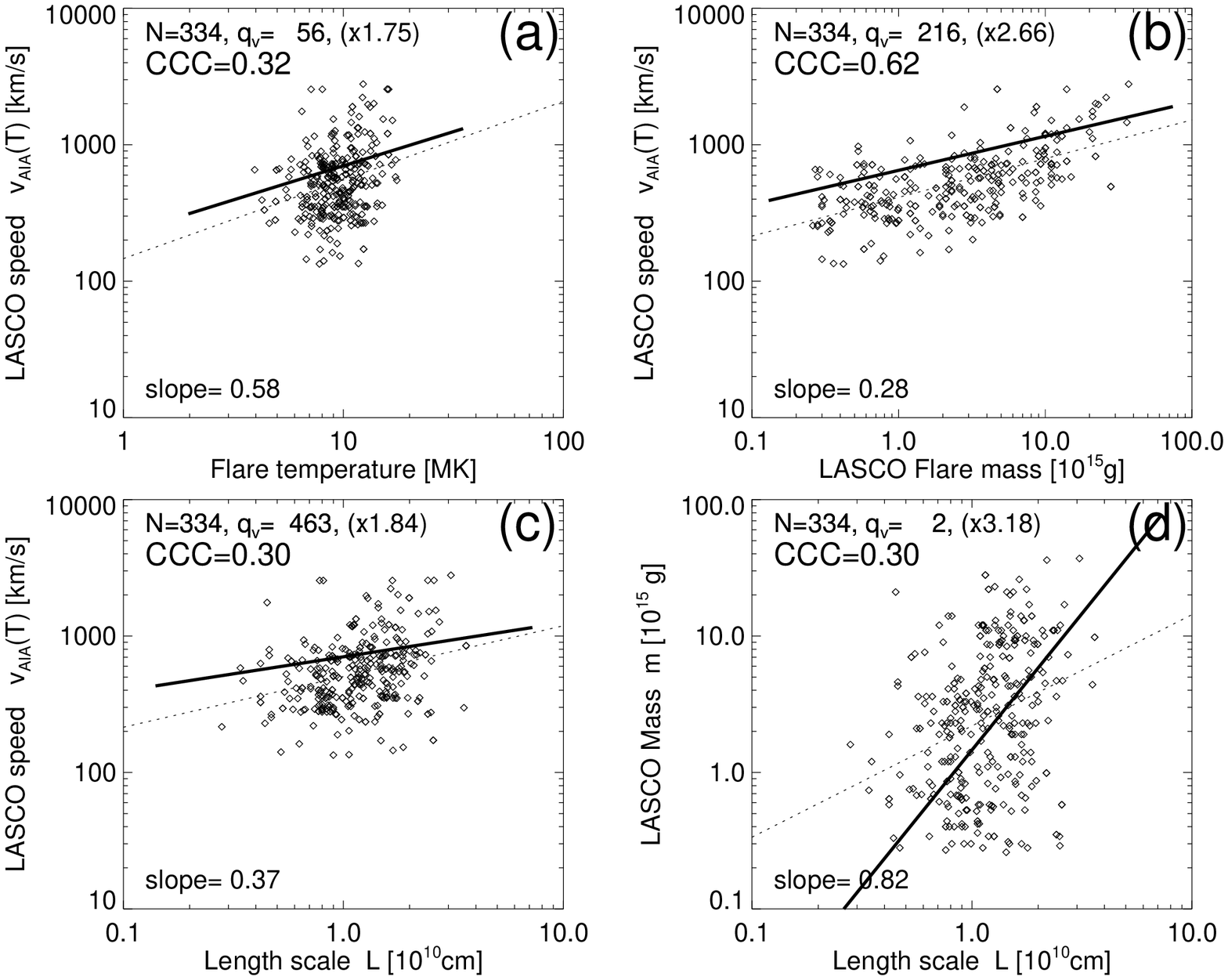}
\caption{Predicted scaling laws tested with LASCO data. 
The datapoints of joint LASCO and AIA events ($N=334$) are
marked with diamonds, a linear regression fit with a dotted
line, and the theoretical scaling law prediction with a
a solid black line (predicted in absolute terms without
adjustment).}
\end{figure}

\begin{figure}
\plotone{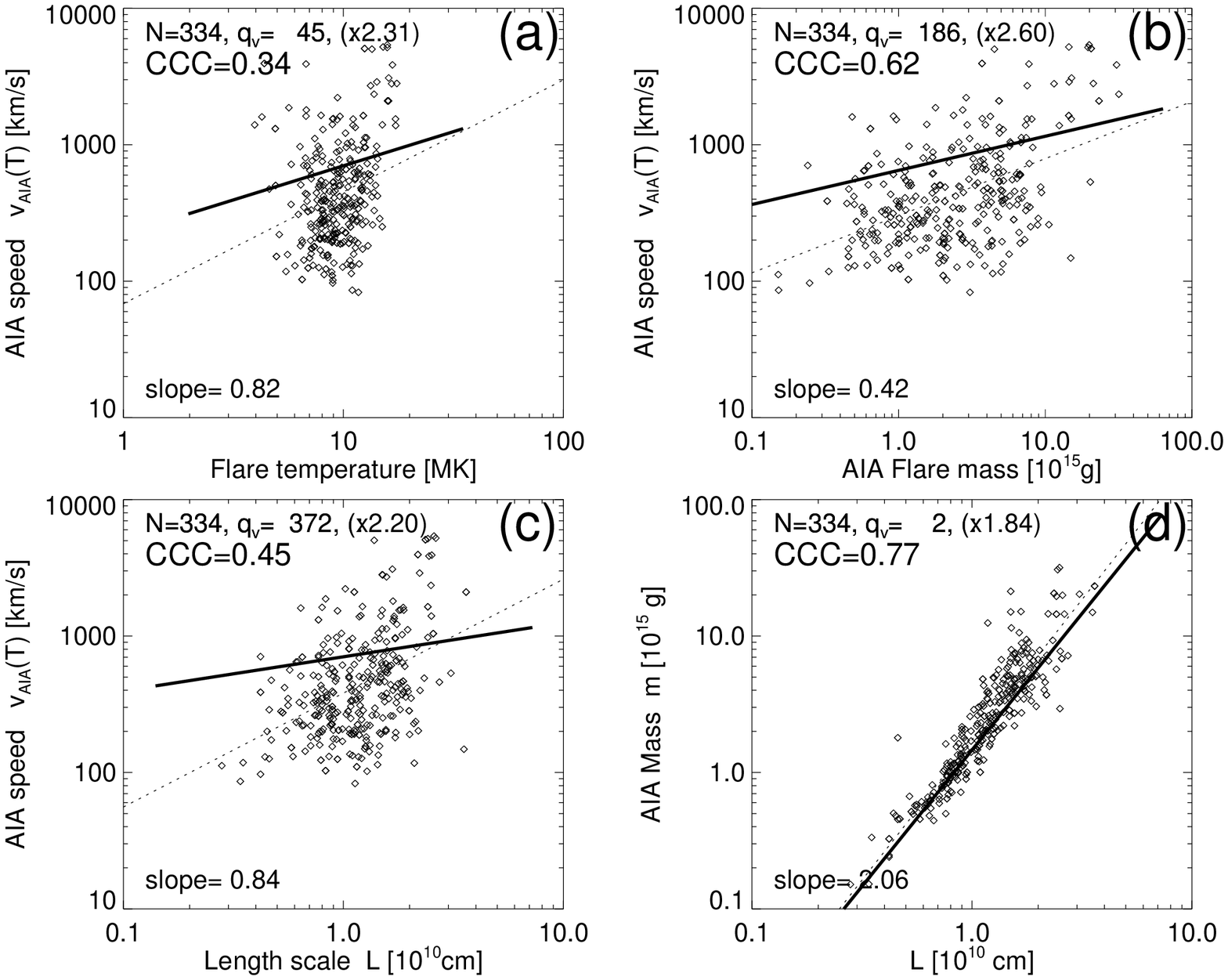}
\caption{Predicted scaling laws tested with AIA data.
The datapoints of joint LASCO and AIA events ($N=334$) are
marked with diamonds, a linear regression fit with a dotted
line, and the theoretical scaling law prediction with a
a solid black line (predicted in absolute terms without
adjustment).}
\end{figure}

\begin{figure}
\plotone{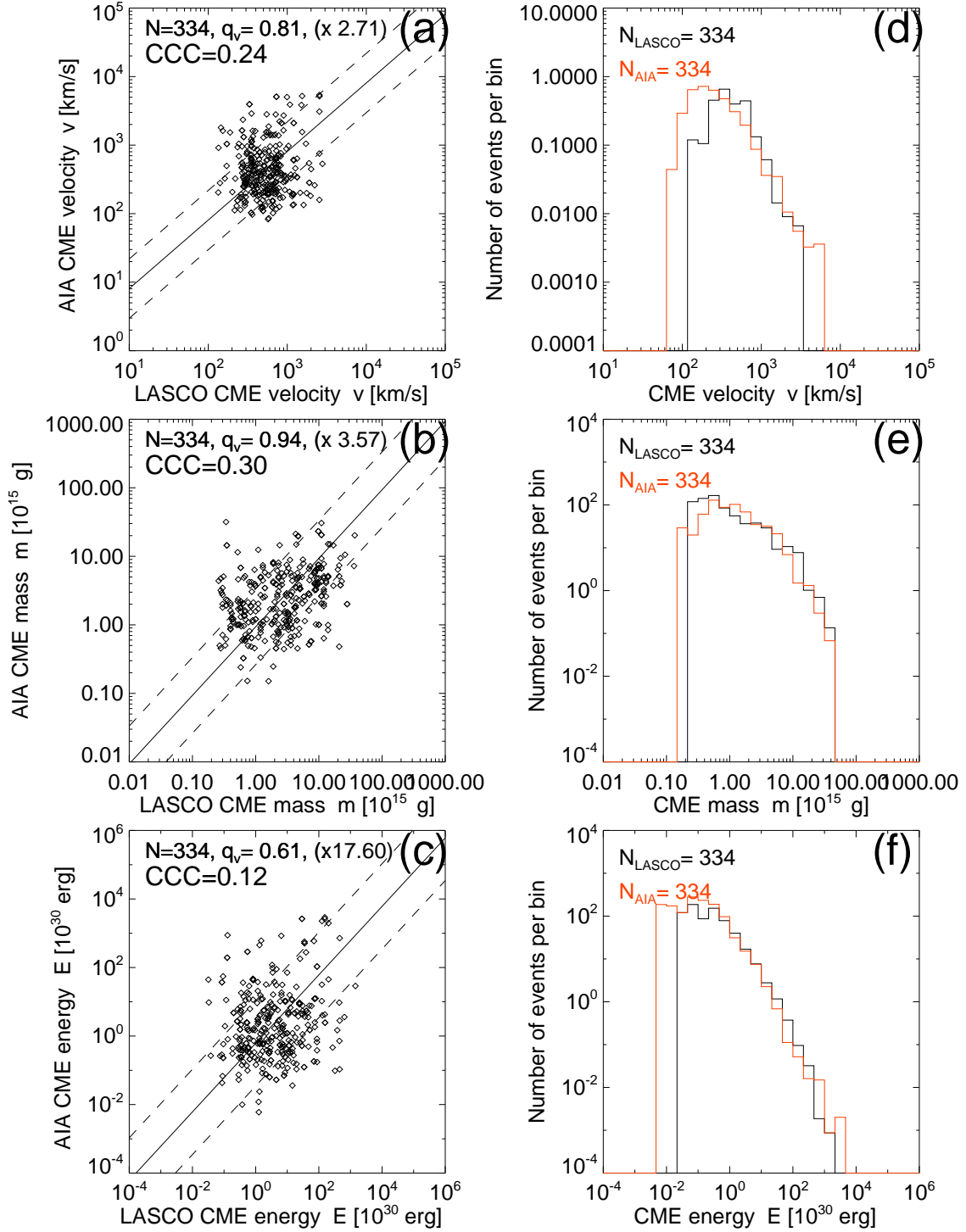}
\caption{AIA-inferred versus LASCO-inferred parameters:
(a) CME velocity, (b) CME mass, and (c) CME energy. The
ratios $q_v$ are obtained from the logarithmic averages 
(indicated with a solid line), and the standard deviation
factors are indicated with dashed lines. The corresponding
size distributions (d,e,f) are displayed with a black
histogram for the LASCO events, and with a red histogram
for the AIA events.}
\end{figure}

\end{document}